\documentclass{article}
\usepackage[numbers]{natbib}
\usepackage[a4paper,margin=1in]{geometry}

\usepackage[utf8]{inputenc}
\usepackage[T1]{fontenc}
\usepackage{mathptmx}
\usepackage{etoolbox}
\usepackage{graphicx}
\usepackage{multirow}%
\usepackage{amsmath,amssymb,amsfonts}%
\usepackage{amsthm}%
\usepackage{mathrsfs}%
\usepackage{dcolumn}
\usepackage{xcolor}
\usepackage{booktabs}%
\usepackage{algorithm}%
\usepackage{algorithmicx}%
\usepackage{algpseudocode}%
\usepackage{listings}%
\usepackage{enumitem}
\usepackage{relsize}
\usepackage{hyperref}
\usepackage{lineno}
\hypersetup{
    colorlinks=true,
    linkcolor=blue,
    filecolor=magenta,      
    urlcolor=blue,
    citecolor=blue,
    }
\usepackage{siunitx}
\newcommand{\air}{\mathrm{air}}
\newcommand{\lamp}{\mathrm{lamp}}
\newcommand{\capsid}{\mathrm{cap}}
\newcommand{\spin}{\mathrm{spin}}

\makeatletter
\def\@email#1#2{%
 \endgroup
 \patchcmd{\titleblock@produce}
  {\frontmatter@RRAPformat}
  {\frontmatter@RRAPformat{\produce@RRAP{*#1\href{mailto:#2}{#2}}}\frontmatter@RRAPformat}
  {}{}
}%
\makeatother

\begin{document}


\begin{center}

\textbf{Computational Fluid Dynamics and Discrete Dipole Approximation Applied to Predict Airborne Viruses Disinfection by Ultraviolet Irradiation}

\vspace{2ex}

\textbf{Talib Dbouk}$^{1*}$ \textbf{and Maxim Yurkin}$^{1}$

\vspace{2ex}

$^1$CNRS, CORIA, UMR 6614, University of Rouen Normandy, Rouen FR-76000, France

\vspace{2ex}

$^*$Corresponding author: talib.dbouk@coria.fr
\end{center}


\begin{abstract}
Ultraviolet (UV) inactivation of airborne viruses involves coupled multiphysics and multiscale processes, yet UV scattering within virus-laden saliva droplets is commonly neglected. Here, for the first time we combine computational fluid dynamics (CFD) and the discrete dipole approximation (DDA) to investigate the influence of droplet-scale UV scattering on airborne virus inactivation. A benchmark CFD--DDA case study is developed and demonstrated that infected saliva droplets can exhibit highly non-uniform UV scattering under representative flow and irradiance conditions. This non-uniformity can substantially affect the predicted number of surviving viral copies, $N_s$, indicating that UV scattering should be considered when modeling UV-based airborne disinfection systems. We systematically investigate the effect on $N_s$ by varying: the initially emitted saliva droplets diameter distribution, their number ($n(t=0)=10^3$ or $10^4$), volumetric airflow rate ($Q_v=100$, 150, or \qty{200}{\m\cubed\per\hour}), droplet--wall interaction (\textit{stick} vs \textit{escape}), and coronavirus UV susceptibility constant ($Z=0.018$ or \qty{0.377}{\m\squared\per\J}). The simulations consider initially infected saliva droplets containing \num{7.4e8} RNA copies/mL, with a viral capsid diameter of $d_\capsid=\qty{100}{\nm}$, at an air temperature of \qty{21}{\degreeCelsius} and relative humidity of 65\%. UV irradiation is represented by a \qty{10}{\W} cylindrical UV-lamp (\qty{34.3}{\mW\per\cm\squared}). The DDA results further enable development of an extended virus-inactivation model based on the classical Chick--Watson law that accounts for UV scattering in both spherical and nonspherical infected droplets. The results highlight the importance of resolving local UV--droplet interactions in airborne virus disinfection models and demonstrate the potential of the coupled CFD--DDA approach for improving the design and performance of UV-based airborne disinfection systems.

\vspace{2ex}

\textbf{keywords}: light scattering, germicidal irradiation, UV-C air purifiers, virus disinfection, saliva, aerosols, computational fluid dynamics, discrete dipole approximation
\end{abstract}



\section{Introduction} \label{sec:intro}

The last COVID-19 coronavirus pandemic \cite{Seitz2020, McNeely2021, Dbouk2021b} was a huge worldwide social challenge. It induced devastating socio-economic impacts which shed light on the importance of protecting the human populations against future transmissions of deadly airborne viruses. This is also affected by different environmental conditions indoors \cite{Dbouk2021a} and outdoors \cite{Dbouk2020a, Dbouk2020weather, Dbouk2021c}. In the years following the COVID-19 pandemic, researchers have been focusing their efforts on the fabrication of new vaccines, while others have been investigating the role of social distancing \cite{Kramer2021}. 

\begin{figure}[ht]
    \includegraphics[width=\textwidth]{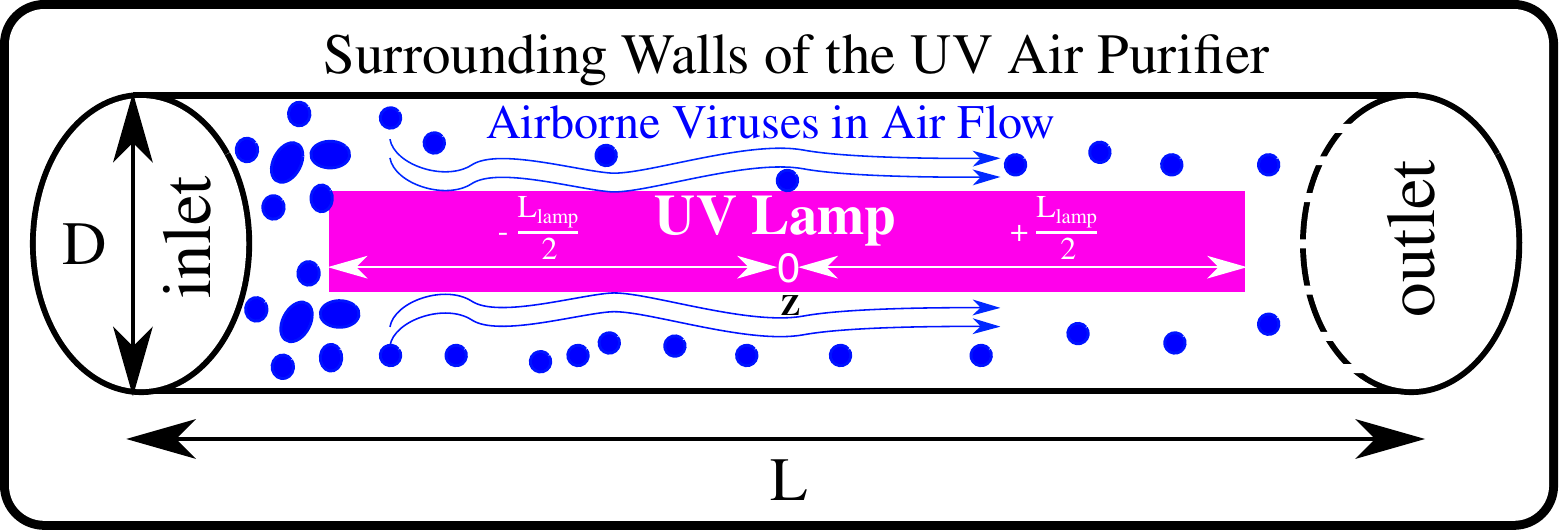}
    \caption{Benchmark design of ultraviolet irradiation air purifier for inactivation of airborne viruses. $L/D=5.625$ and $L=0.765~m$.}
    \label{fig:BenchMarkDesign}
\end{figure}

Very few scientists have been attempting to propose reliable engineering solutions, such as ultraviolet (UV) air purifiers, that can inactivate airborne viruses \cite{Trivellin2021}. Studies on the numerical modeling of such \textit{airborne} virus disinfection goes back to the early work by Alic \cite{Alic2021ThreeStageInactivation}. The author developed an analytical model with a 3-stage inactivation of viruses within droplets and solid porous particles.
To better understand the mechanisms underlying the complex multiscale multiphysics phenomena involved, they defined a simple benchmark design/geometry (figure 3 in Ref. \cite{Alic2021ThreeStageInactivation}) of UV air purifier as shown in figure \ref{fig:BenchMarkDesign}. This geometry is adopted in the present work with $L/D=5.625$ ($L=0.765~m$) and UV lamp of $L_\lamp=\qty{46.5}{\cm}$, $D_\lamp=\qty{2}{\cm}$ and $P_\mathrm{UV}=\qty{10}{\W}$ as a benchmark computational-fluid-dynamics (CFD) study case.

\begin{figure}[ht]
    \includegraphics[width=\textwidth]{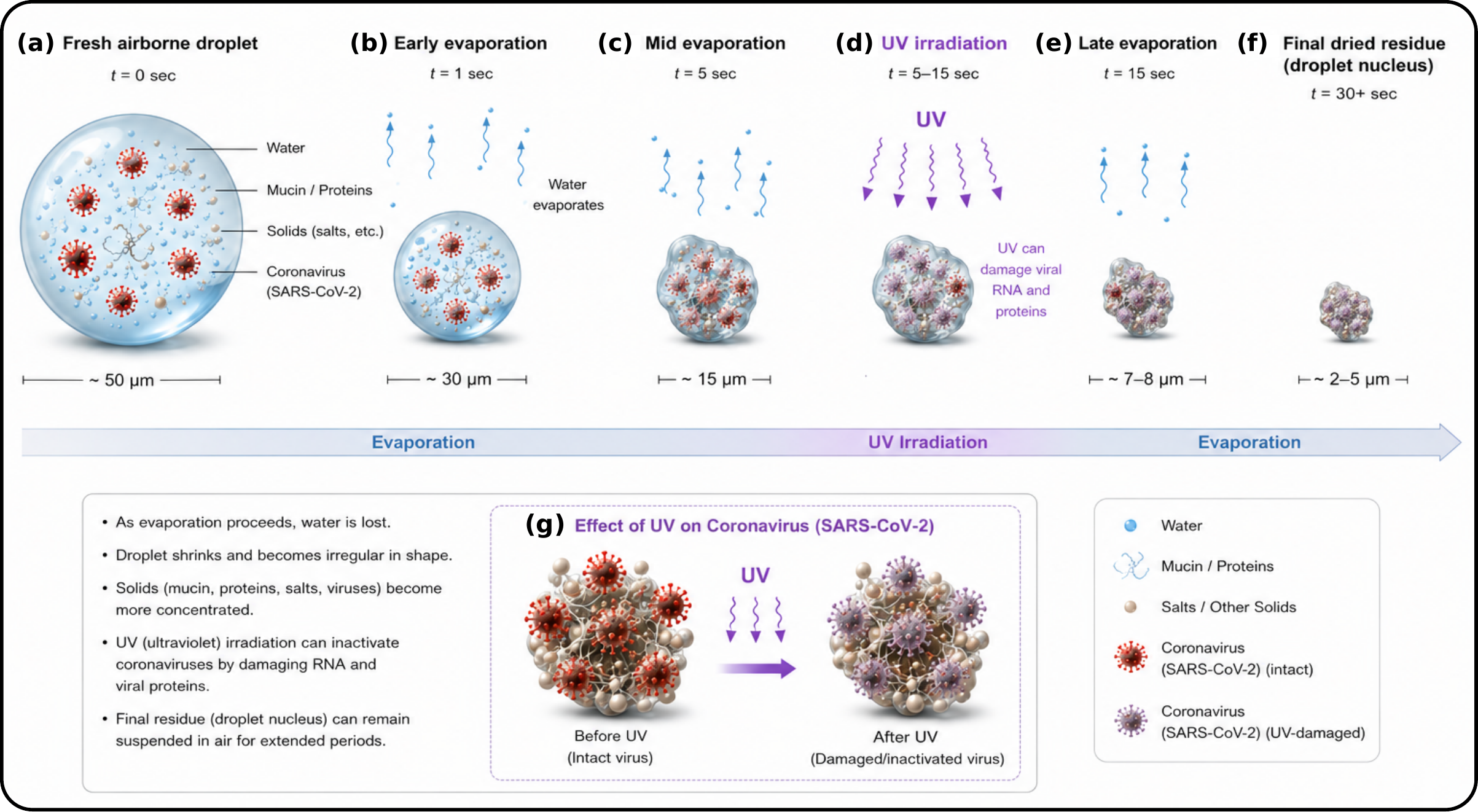}
    \caption{Effect of evaporation on altering the regular shape of the respiratory droplet. \textit{Virus Images: AI-assisted Author generated}.}
    \label{fig:Evaporation_COV_UV}
\end{figure}

Research studies on \textit{airborne} UV germicidal irradiation (UVGI) employing CFD are rare in the literature. For example, Marshall et al. \cite{Marshall2024} and Sankurantripati et al. \cite{SANKURANTRIPATI2025} are the only authors, to our knowledge, who developed CFD to investigate airborne virus inactivation using UV air purifier design applied to automobile and transportation (bus) industry. With large eddy simulation of the air flow and Euler-Lagrange formulation with two-way coupling and evaporation model, the authors proposed a new engineering equipment design and made attempts to quantify its efficiency in inactivating airborne virus. They, however, made strong assumptions in their studies. First, they considered quasi uniform size of droplets that are only treated as spheres with initial diameters between 2 and \qty{4}{\um}. 
By contrast, recent experiments by Sapkota et al. \cite{Sapkota2026} showed that airborne saliva droplets can obtain very irregular shape due to evaporation under specific conditions of temperature, relative humidity (RH), air flow speed, and organic composition, as shown schematically in figure \ref{fig:Evaporation_COV_UV}. Moreover, the size distribution of respiratory emitted saliva aerosol can be very wide and multi-modal, e.g., from \qty{1}{\um} to more than \qty{500}{\um}, depending on the activity of mouth and vocal chords in addition to the health condition of a person \cite{Xie:2009,Han2013,Asadi2020VoicingAerosol}. Second, the UV intensity has always been assumed uniform in the droplets, which is generally not the case.

Experimental measurements of airborne viruses disinfection by UV irradiation are rare as well \cite{Jensen1964,Walker2007,Bedell2016,Welch2018}. Each experiment included a completely different engineering system design with a focus on the air flow circulation in an indoor space that directs infected airborne droplets (aerosols) close to the UV source (lamp or LED), in order to increase the droplets exposure to UV irradiation. There is unfortunately no clear proof that the UV field in these devices is the main factor responsible for the inactivation of the airborne viruses. This omission stems from the true complexity of conducting such measurements due to the multiscale and multiphysics nature of the dispersed media. Consider tiny dispersed  droplets with complex flow dynamics on sub-meter scales dependent on specific engineering system design and several physical processes, like evaporation, coalescence, shearing, and interaction of airborne droplets with the surrounding interfaces, as reviewed recently by Longest et al. \cite{Longest2024}. 

This means that UV susceptibility of airborne viruses is extremely difficult to quantify through aerosol in UV field experiments. To address this problem, we develop an innovative platform combining CFD with discrete dipole approximation (DDA) in order to deepen our understanding of the mechanisms responsible for the inactivation of airborne viruses in-saliva droplets induced by ultraviolet irradiation. For the CFD part, we consider two different size distributions of saliva droplets, extending previous works \cite{Marshall2024,SANKURANTRIPATI2025}. DDA is a well-established numerical method for calculating scattering and absorption of electromagnetic waves by particles of arbitrary shape and composition \cite{Yurkin2007DDAReview,Yurkin2023Chapter}. We apply it to obtain distribution of UV intensity inside a single saliva droplet, which allows us, for the first time, to account for variation of obtained UV dose with droplet shape (i.e., induced by evaporation) and specific virus position inside it.

Finally, we relate our simulation results to the design of engineering systems intended for disinfection of airborne viruses by UV irradiation.

\section{Investigated Parameters}

In the present work, we investigate the effect of: 
\begin{itemize}[label=\checkmark]
    \item Different initial size distributions of effective saliva droplets diameters: ($\qty{1}{\um} \leq d \leq \qty{500}{\um} $ and $\qty{1}{\um} \leq d \leq \qty{20}{\um}$ for evaporative droplets, and $\qty{1}{\um} \leq d \leq \qty{5}{\um}$ for non-evaporative dropelt, e.g. dry saliva droplets), 
    \item Initial number of infected saliva droplets in the system ($10^3$ or $10^4$ droplets),
    \item Input air flow rate (100, 150, or \qty{200}{\m\cubed\per\hour}),
    \item Droplet-wall interaction models (\textit{stick} versus \textit{escape}).
    \item UV wavelength within UV-C range ($\lambda=\qty{254}{\nm}$ or $\lambda=\qty{280}{\nm}$),
    \item Virus UV susceptibility ($Z=0.0134$ or \qty{0.377}{\m\squared\per\J}),
    \item UV scattering versus neglected UV scattering in airborne infected dried saliva droplets.
\end{itemize}
on the total number of final survived copies $N_s$ of a coronavirus (virion capsid diameter of \qty{100}{\nm} \cite{tyrrell1996coronaviruses}) in air circulation benchmark system/design, operating at \qty{21}{\degreeCelsius} and 65\% relative humidity (RH); see figure \ref{fig:BenchMarkDesign}.

\subsection{Initial size distribution}\label{sec:sizeDistribution}

Concentration and size distribution of saliva droplets emitted from a person mouth to the surrounding air, depends on many factors, such as the anatomy, health condition of a person, and the level of mouth activity: e.g., speaking calmly, speaking loudly, shouting, coughing, or sneezing \cite{Xie:2009}. For instance, speaking loudly can generate up to thousands of saliva droplets per second \cite{Xie:2009,Asadi2020VoicingAerosol,Stadnytskyi2020AirborneLifetime}. 
Han et al. \cite{Han2013}, who focused on sneezing, found that depending on subjects both uni-modal (one peak exist) and bi-modal (two peaks exist denoted peak 1 and peak 2) distributions exist. For example, they reported that in the case of a uni-modal distribution, the geometric mean of droplet size of all the sneezes is about 360.1 $\mu$m. However, in the case of bi-modal distribution, the geometric mean is 386.2 $\mu$m for peak 1 and 72.0 $\mu$m for peak 2. Note also that large saliva droplets ($>\qty{100}{\um}$) are fewer in number and quickly fall in air flow, while more numerous smaller droplets ($<\qty{100}{\um}$) remain suspended for longer periods of time and are, thus, more important for infection transmission.

To cover most of the possible scenarios of saliva emission rates, we fit the literature data with the Rosin--Rammler distribution (also called the Weibull particle size distribution \cite{Weibull:1951}) given by the following probability density function (PDF):
\begin{equation}\label{eq:PDF}
    f(d)=
    \begin{cases}
        \dfrac{
            (n/\beta)(d/\beta)^{n-1}
            \exp\!\left[-(d/\beta)^n\right]
        }
        {
            \exp\!\left[-(d_{\min}/\beta)^n\right]
            -
            \exp\!\left[-(d_{\max}/\beta)^n\right]
        },
        &
        d_{\min}\le d \le d_{\max},\\[2ex]
        0, & \text{otherwise.}
    \end{cases}
\end{equation}

\begin{figure}[ht]
    \includegraphics[width=\textwidth]{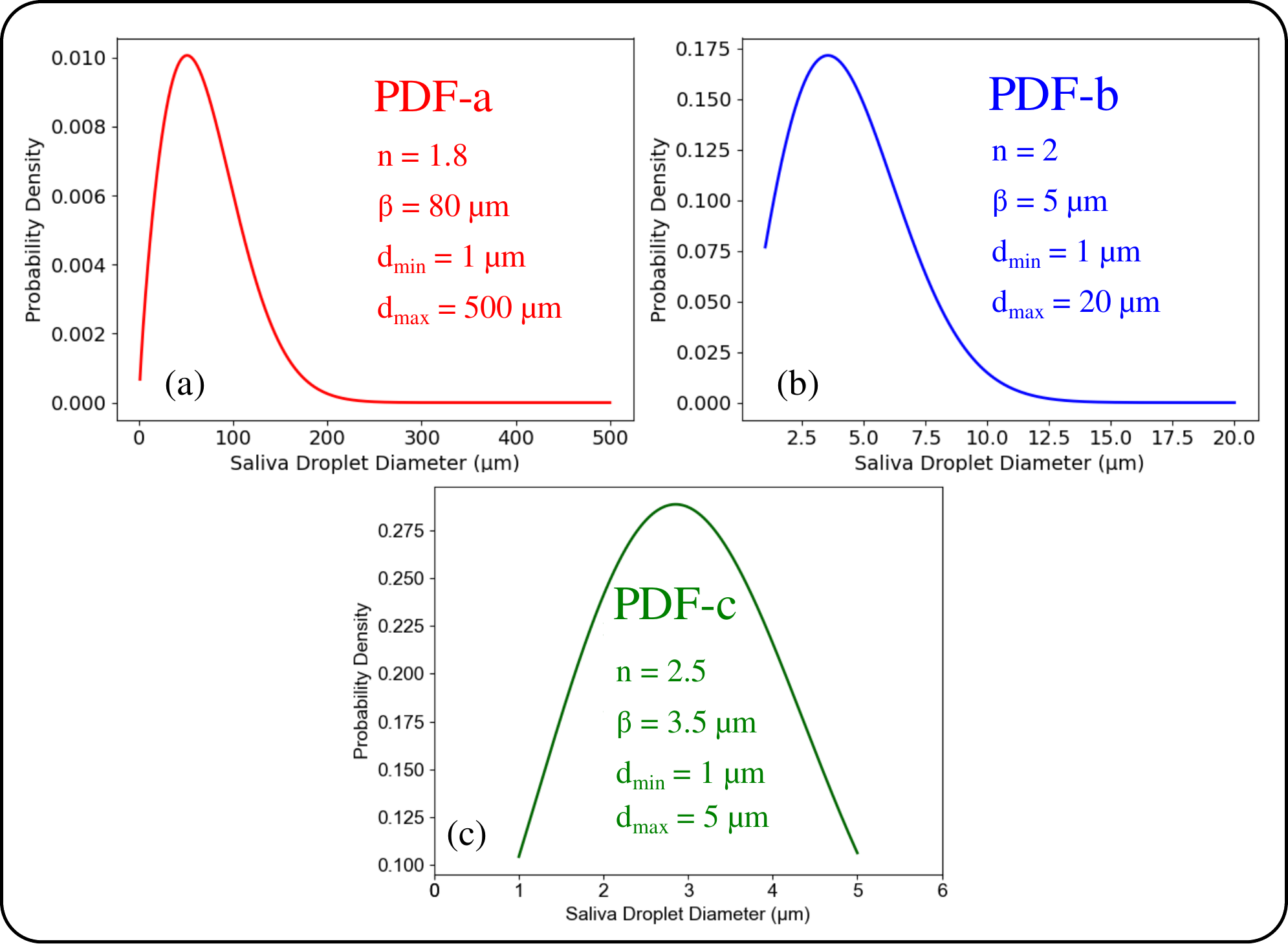}
    \caption{Probability density function (PDF) example of two different size distributions of emitted saliva droplets. (a) PDF-a (dehydrated); (b) PDF-b (quasi dehydrated); (c) PDF-c (dry). PDF-a, PDF-b and PDF-c each follow a Rosin--Rammler (Weibull) distribution function with different ($n$, $\beta$, $d_{min}$, $d_{max}$ parameters, see equation \eqref{eq:PDF}.}
    \label{fig:PDF}
\end{figure}

Two integral characteristics of the distribution are the number mean diameter $D_{10}$ (standard mean of the PDF) and the Sauter mean diameter $D_{32}$, equal to the ratio of third and second moments of the PDF. In this paper, we consider three PDF's shown in figure \ref{fig:PDF}, denoted further as PDF-a, PDF-b for evaporative droplets and PDF-c for dry non-evaporative saliva droplets. Moreover, the initial number of droplets at $t=0$ investigated is $10^3$ and $10^4$.

\subsection{Saliva composition and initial viral load}\label{sec:viralLoad}

Apart from carrying airborne viruses, saliva plays an important role in maintaining oral health and supporting digestion. It is mostly water (about 99.5\%), but it also contains enzymes, proteins, minerals, and antibodies. The average adult person produces about 0.5 to 1.5 liters of saliva per day. The composition of human saliva\cite{Chiappin2007,Strickleton2014,Zhang2022,Jakubovics2014,Sanchez2013, humphrey2001role,carpenter2013salivary} depend on each person and the type of his/her activity like speaking or eating. In other words, it depends on the flow-rate of saliva secretion by the major glands. An example of a general composition\cite{Chiappin2007,Strickleton2014,Zhang2022} of human saliva is given in table \ref{tab:salivaComposition}.


\begin{table}[ht]
\centering
\begin{tabular}{lll}
\toprule
\textbf{Target} & \textbf{Size} & \textbf{Concentration} \\
\midrule
Dissolved salts & $<\qty{1}{\nm}$ & \qtyrange{0.2}{0.5}{\mg\per\mL} \\ 
Proteins & \qtyrange{2}{10}{\nm} & \qtyrange{0.5}{1}{\mg\per\mL} \\
Mucin aggregates & \qtyrange{10}{100}{\nm} & \qtyrange{0.1}{0.4}{\mg\per\mL} \\
Bacteria & \qtyrange{500}{900}{\nm} & \qtyrange{0.01}{0.1}{\mg\per\mL} \\
\bottomrule
\end{tabular}
\caption{Composition of human saliva produced during speech.}\label{tab:salivaComposition}
\end{table}

Wyllie et al. \cite{Wyllie2020} measured the viral loads of SARS\nobreakdash-CoV\nobreakdash-2 in saliva of different infected patients to be up to \num{3e10} RNA copies/mL, making it more sensitive for SARS\nobreakdash-CoV\nobreakdash-2 detection than nasopharyngeal swabs. However, for most patients the viral load was within \num{7.4e8} copies-RNA/mL, this value is further used as characteristic viral load during the acute infection phase. Assuming a single RNA per virion, the initial number of virions $N_{i,0}$ (at $t=0$) in each saliva droplet is proportional to its volume:

\begin{equation}\label{eq:N0}
    N_{i,0} = \num{7.4e-4} \frac{\pi}{6} \left(\frac{d_i}{\unit{\um}}\right)^3
\end{equation}
where $d_i$ is the initial diameter of $i$-th saliva droplet, that is set according to the one of the initial PDF distribution function shown in figure \ref{fig:PDF}. 

We assume that evaporation process only removes water content from the infected saliva droplets, while all the virions remain, increasing their concentration. Moreover, we neglect coalescence and breakup phenomena due to the relatively small number of initially emitted saliva droplets (and, thus, low volume fraction) in the surrounding air shown in figure \ref{fig:BenchMarkDesign}.

\subsection{Optical model for UV light scattering}\label{sec:optmodel}

Although saliva is definitely inhomogeneous (see table \ref{tab:salivaComposition}), its components (including virions) are either much smaller than the considered UV wavelength ($\lambda=254$ or \qty{280}{\nm}) or have very low concentration (as bacteria). Moreover, the largest optical contrast (difference of refractive indices) is between water and all other components. Thus, in a dried state, the contrast between individual components is much smaller. Therefore, we consider saliva droplets as homogeneous, so its optical properties are fully determined by its size, shape, and complex refractive index $m$. The latter effectively accounts for all the constituents.

No experimental measurements of $m$ in UV range exist yet for either SARS\nobreakdash-CoV\nobreakdash-2 virions or for saliva droplets. However, the SARS\nobreakdash-CoV\nobreakdash-2 virions were previously modeled as a mixture of protein, RNA, lipids, tryptophan, tyrosine, phenylalanine, and cystine, resulting in $m=\complexnum{1.59+0.0017i}$ at \qty{260}{\nm} \cite{Doughty2021}. More recently, the same authors \cite{Hill2021, Hill2024} calculated refractive indices at \qty{254}{\nm} for both coronavirus and saliva -- $m=\complexnum{1.61+0.0014i}$ and $m=\complexnum{1.57+0.006i}$, respectively. In this paper, we assumed round values of $m=\complexnum{1.6+0.001i}$ for the DDA simulations, which corresponds to the highest possible $\operatorname{Re}m$ (corresponding to dried saliva or proteins) and relatively low value of $\operatorname{Im}m$. However, the latter should not lead to any significant differences for simulated droplet sizes of up to \qty{5}{\um}.

\subsection{Saliva droplet evaporation time scale order of magnitude}

For the evaporation time scale analysis, according to Sherwood--Spalding Ranz-Marshall correlations \cite{Ranz:52a,Ranz:52b} for a sphere, a saliva droplet in air flow at \qty{3}{\m\per\s} (present benchmark design at \qty{100}{\m\cubed\per\hour}, $T_\air=\qty{21}{\degreeCelsius}$, $\mathrm{RH}=65\%$) undergoes diameter reduction from \qty{10}{\um} to \qty{1}{\um} in about \qty{1}{\s}, which is larger than the residence time scale at this flow rate; see figure \ref{fig:U_UVC_P}. Moreover, this rough estimate of \qty{1}{\s} neglects the transformation of saliva droplet from liquid to gel-like particle during evaporation due to protein, salts, and mucin residues. This transition will in fact induce much longer times of evaporation; e.g. see Sapkota et al. \cite{Sapkota2026} and figure \ref{fig:Evaporation_COV_UV}.

\section{Air circulation in presence of UV field}

\subsection{Fluid flow circulation benchmark design}

In order to decouple the complex multiscale multiphysics phenomena involved in the UV disinfection of airborne viruses in dynamic infected saliva droplets, a CFD 3D benchmark design is developed. It consists of a cylindrical tube responsible for air flow circulation. The air passing through the tube is thus a carrier/circulator of infected airborne saliva droplets. A UV cylindrical lamp is installed at the central position of the tube which will allow the exposure of infected saliva droplets to UV irradiation with different doses, of course depending on multiple conditions that will be developed and explained in details.

\subsection{UV field modeling within CFD}

UV-C irradiance flux of about \qty{34.3}{\mW\per\cm\squared} is considered as a very high UV power case scenario. For example, Kariwa et al. \cite{Kariwa2004}, said: "Heating the virus at \qty{56}{\degreeCelsius} for \qty{5}{\minute} dramatically reduced the infectivity of the virus from \num{2.6e7} to 40~TCID$_{50}$/mL, whereas heating the virus for \qty{60}{\minute} or longer eliminated all infectivity. Irradiation with ultraviolet light at \qty{0.134}{\mW\per\cm\squared} for \qty{15}{\minute} reduced the infectivity from \num{3.8e7} to 180~TCID$_{50}$/mL; However, prolonged irradiation (\qty{60}{\minute}) failed to eliminate the remaining virus, leaving 18.8~TCID$_{50}$/mL." 

In the present CFD simulations, a UV lamp is considered such that:  $R_{Lamp}=\qty{0.01}{\m}$, $L_{Lamp}=\qty{0.465}{\m}$, $P_\mathrm{UV}=\qty{10}{\W}$, thus power density of ${P_\mathrm{UV}}/(2 \pi RL) = \qty{34.3}{\mW\per\cm\squared}$. This is 256 times larger than the irradiance flux applied by Kariwa et al. experiments \cite{Kariwa2004}.

In specific UV-C lamp air purifier designs with air flow, under specific conditions it will be indeed the lamps high temperature in direct contact with touching droplets,  the one responsible for virus inactivation (guided of course by the local air flow), and thus not always the UV-C light scattering or accumulated dose the one responsible.
In the literature, the community never paid attention to the fact of "complex multiphysics coupling" in this problem, and they always made "indirect conclusions"; e.g. by looking only to one single path, without quantifying which "part of physics" or "contributor" is indeed the one that is the most responsible for inactivating the airborne viruses.
It is worth noting that this is an extremely difficult information to extract experimentally, if not impossible, and thus this explains the present adopted advanced multiphysics multiscale modeling CFD-DDA approach.

We will try to show that when light scattering is not uniform locally in infected saliva droplets then the UV disinfection system including: infected saliva droplet irregular shape, e.g. due to evaporation, in addition to flow conditions, the design and the scale, each can differently impact the UV susceptibility of virus.

\section{Computational fluid dynamics modeling}

\subsection{CFD modeling of fluid-particles coupling}

A transient Eulerian--Lagrangian solver is developed for compressible turbulent flows with Lagrangian particles (droplets with virus load inclusions number).

The continuous air phase is solved using the Finite-Volume Method \cite{Moukalled2015}, while parcels are tracked individually in a Lagrangian framework. The air (as multi-species mixture to account of humidity) and particle phases can thus exchange: mass, momentum, energy and species. Pressure--velocity coupling is achieved through the PIMPLE algorithm \cite{Jasak2009}. In the present work, fluid-particles two-way coupling is applied in addition to second-order schemes in time and space for the CFD simulations. For the flow field, grid sensitivity analysis is developed and a 3D hexahedral mesh with local refinement at the walls is adopted following the Grid Convergence Index (GCI) approach \cite{roache1994perspective,Celik:2008}.

It is important to remind that in fluid-particle coupling, the "one-way" coupled approach means that each droplet sees the original air flow velocity field, and that air is unchanged regardless of how many droplets are present. While in the "two-way" coupling approach each droplet experiences drag, and the equal-and-opposite drag force is added to the air momentum equation. As the air flow slows down, accelerates, heats up, cools down, or changes composition, subsequent droplets will move through a modified air flow field. Thus, droplets trajectories become indirectly coupled through the continuous phase, even though there are still no direct particle-particle forces. This latter effect becomes even more important at moderate to high droplets loadings or droplets clouds where momentum, heat, or mass exchange significantly alters the carrier gas phase (e.g. air in this case).


\subsection{Air phase governing equations}

\subsubsection{Continuity equation}

The air mixture continuity equation is given by:

\begin{equation}
\frac{\partial \rho}{\partial t}
+\nabla\cdot(\rho\mathbf{U})
=
0
\end{equation}
where $\rho$ is the gas density, $\mathbf U$ the gas velocity and $S_m$ is the mass source due to particle evaporation or reactions. 


\subsubsection{Momentum equation}

The compressible Navier--Stokes equation is given by:

\begin{equation}
\frac{\partial (\rho\mathbf U)}{\partial t}
+
\nabla\cdot(\rho\mathbf U\mathbf U)
=
-\nabla p
+
\nabla\cdot\boldsymbol{\tau}
+
\rho\mathbf g
+
\mathbf S_p,
\end{equation}
where the viscous stress tensor is given by:

\begin{equation}
\boldsymbol{\tau}
=
\mu_{\rm eff}
\left[
\nabla\mathbf U
+
(\nabla\mathbf U)^T
-
\frac23(\nabla\cdot\mathbf U)\mathbf I
\right].
\end{equation}

The momentum source term $\mathbf S_p$ accounts-to/results-from interphase momentum exchange.


\subsection{\texorpdfstring{$k$--$\epsilon$}{k-epsilon} turbulence modeling}

The standard $k$--$\epsilon$ model is adopted in the present work. It is one of the most widely used Reynolds-averaged Navier--Stokes (RANS) turbulence models in CFD simulations. It falls within the family of 2-equations eddy-viscosity models where turbulence is taken into account by transport equations in: turbulent kinetic energy (denoted $k$) and its dissipation rate (denoted $\epsilon$).
In the present platform, a classical Launder--Spalding formulation together with an optional Rapid Distortion Theory (RDT) compression correction for compressible flows are considered. This make the turbulence model suitable for both incompressible and compressible simulations and requires appropriate near-wall treatment through wall functions or low-Reynolds-number formulations. For the detailed mathematical formulation of the $k-\epsilon$ turbulence model, the readers may refer to the book by Launder and Spalding 1974 \cite{Launder1974}.

\subsubsection{Species transport}

For each $j$-th species in the air mixture ($O_2$, $N_2$, ${H_2}O$), we have:

\begin{equation}
\frac{\partial (\rho Y_j)}{\partial t}
+
\nabla\cdot(\rho\mathbf UY_j)
=
\nabla\cdot
\left(
\rho D_{\rm eff}\nabla Y_j
\right)
+
S_{Y_i},
\end{equation}
where $Y_j$ is the species mass fraction and $S_{Y_i}$ is the parcel's source term. The adopted main typical parcel contribution is evaporation of water content in saliva droplets such that:

\begin{equation}
S_{Y_i}
=
\dot m_{\mathrm{evap},i}
\end{equation}


\subsubsection{Energy equation}

The sensible enthalpy equation is solved as the following:

\begin{equation}
\frac{\partial (\rho h)}{\partial t}
+
\nabla\cdot(\rho\mathbf Uh)
=
\frac{Dp}{Dt}
+
\nabla\cdot
(\alpha_{\rm eff}\nabla h)
+
S_h
\end{equation}
where the particle energy source term due to convection and evaporation is given by:

\begin{equation}
S_h
=
Q_\mathrm{conv}
+
Q_\mathrm{evap}
\end{equation}


\subsubsection{Equation of state}

The air mixture is considered as an ideal gas such that:

\begin{equation}
p=\rho RT
\end{equation}


\subsection{Lagrangian particles governing equations}

Due to the low volume fraction of droplets in a computational fluid element cell, the parcel method is adopted such that each computational parcel can represent single or many identical physical particles inside a fluid control volume.


\subsubsection{Particles position}

A droplet position results from its velocity thanks to the Newton's second law of motion such that:

\begin{equation}
\frac{d\mathbf x_p}{dt}
=
\mathbf U_p
\end{equation}
where
\begin{equation}\label{eq:dUp}
m_p
\frac{d\mathbf U_p}{dt}
=
\mathbf F_D
+
\mathbf F_g
+
\mathbf F_{\rm other}
\end{equation}
where $F_g$ is the gravitational force.
Assuming small size droplets, a common sphere drag coefficient force model can be employed such that:
\begin{equation}\label{eq:F_D1}
\mathbf F_D
=
\frac{m_p}{\tau_p}
(\mathbf U-\mathbf U_p)
\end{equation}
where $m_p$ is the droplet's mass, $F_D$ the drag force and $\tau_p$ is the particle relaxation time or response time that measures how quickly a droplet adjusts its velocity to match the surrounding air fluid velocity.

Assuming $i$-th spherical saliva droplet of diameter $d_i$ moving at low Reynolds number (e.g. Stokes flow), the drag force is given by: 
\begin{equation}\label{eq:F_D2}
\mathbf F_D
=
3 \pi \mu d_i
(\mathbf U-\mathbf U_p)
\end{equation}
where $\mu$ is the air mixture effective dynamic viscosity. Substituting equation \eqref{eq:F_D2} in equation \eqref{eq:F_D1} gives the Stokes relaxation time as: $\tau_p={\rho_p {d_i}^2}/18\mu$.

It is worth noting that additional forces, denoted by $\mathbf F_{\rm other}$ in equation \eqref{eq:dUp}, can be added in the present solver to account for different forces if needed like pressure gradient, lift, virtual mass, and/or Brownian motion. These are negligible in the present work due to the type of saliva droplets under investigation (droplets of effective diameters $\geq \qty{1}{\um}$ at low Reynolds numbers at the particles scale).


\subsubsection{Heat transfer at the droplets scale}

The saliva droplets effective temperature based on energy balance at the scale of each droplet satisfies the following:
\begin{equation}
m_p c_p
\frac{dT_d}{dt}
=
hA_i(T-T_d)
-
\dot m_{\mathrm{evap},i} L
\end{equation}
where $hA(T-T_d)$ represents the convective heat transfer (denoted $Q_\mathrm{conv}$), $\dot mL$ represents the latent heat associated with the phase change from evaporation, $A_i$ the surface area of the droplet, $\dot m$ the mass evaporation rate (\unit{\kg\per\s}) and $L$ is the latent heat of vaporization (\unit{\J\per\kg}).

$h$ is the convective heat transfer coefficient given by the following:
\begin{equation}
    Nu = \frac {h d_i} {k}
\end{equation}
where $k$ is the air flow thermal effective conductivity and $Nu$ the Nusselt number.

To describe the evolution of $h$, the Ranz and Marshall (1952) correlation \cite{Ranz:52a, Ranz:52b} is frequently used for spherical droplets such that:
\begin{equation}
    Nu = 2 + 0.6 {Re_p}^{\frac {1}{2}} {Pr}^{\frac {1}{3}}
\end{equation}
where $Re_p$ and $Pr$ are the Reynolds number at the scale of the saliva droplet and the Prandtl number, respectively. Other advanced correlations can be also used to account for advanced transient effects in the Nusselt number equation (see for example, Dbouk and Drikakis \cite{Dbouk2020weather}).


\subsubsection{Droplets mass}

Assuming virions don't leave the saliva droplets, but only water content due to evaporation, then the saliva droplet's effective mass after evaporation evolves according to the following:

\begin{equation}
\frac{dm_p}{dt}
=
-
\dot m_\mathrm{evap}
\end{equation}


\subsection{Eulerian--Lagrangian coupling}

Particle source terms are accumulated over each computational cell such that:
\begin{equation}
S_m
=
\sum_p
\frac{\dot m_p}{V_{\rm cell}}
\end{equation}
and
\begin{equation}
\mathbf S_p
=
-
\sum_p
\frac{\mathbf F_p}{V_{\rm cell}}
\end{equation}
in addition to:
\begin{equation}
S_h
=
\sum_p
\frac{\dot Q_p}{V_{\rm cell}}
\end{equation}

For the species inside the air flow mixture:
\begin{equation}
S_{Y_i}
=
\sum_p
\frac{\dot m_{i,p}}{V_{\rm cell}}
\end{equation}


\subsection{UV CFD-DDA 3D numerical platform}

The CFD and DDA solvers are employed explicitly in the present works, but our long term objective is to implicitly couple the CFD solver to the DDA solver in a unique new numerical platform that can handle implicitly the following tasks at every time step $t_k$:

\begin{itemize}[label=\$]
\item Solve UV Lamps Irradiance fields within the 3D CFD solver,
\item Solve UV light scattering fields within the 3D DDA solver for a specific droplet shape,
\item Update the UV dose accumulated in each infected saliva droplet based on DDA outputs,
\item Update the total number of survived virus copies in each infected saliva droplet in the 3D CFD solver,
\item Move all the saliva droplets,
\item Update the evaporation change of phase/mass,
\item Assemble all the parcels source terms,
\item Solve the air mixture flow continuity,
\item Solve the momentum,
\item Solve the pressure-velocity correction,
\item Solve the species transport,
\item Solve the energy equation,
\item Update the turbulence quantities,
\item Repeat the PIMPLE algorithm corrections until flow convergence,
\item Go up and repeat all tasks for the next time step $t_{k+1}$.
\end{itemize}


\subsection{Physical models in the UV CFD-DDA numerical platform}

The new unique 3D CFD-DDA particles parcel framework will be able to account for different particle's physics like: UV irradiation and UV light scattering, kinematic tracking, drag, turbulent dispersion, heat transfer, evaporation, breakup, collision, coalescence, devolatilization, heterogeneous reactions, char combustion, gas-phase chemistry, radiation and surface-film interaction.

Assuming droplet-to-air very low volume fraction of airborne saliva droplets that are micrometers in size, the following main particles physics models and sub-models are considered in the present study as shown in table \ref{tab:activeModels}. 

\newpage 

\begin{table}[ht]
    \centering
\begin{tabular}{ll}
\toprule
\textbf{Particle Physics} & Models and Sub-Models\\
\midrule
Species & Released species ($H_2O$)\\
Energy & Heat transfer and latent heat\\
Position & Lagrangian tracking\\
Momentum & Newton's second law\\
Forces & Drag and gravity\\
Temperature & Heat balance\\
Mass & Evaporation\\
UV dose & UV view factor\\
UV scattering & DDA method\\
Particle-Gas & two-way coupling\\
\bottomrule
\end{tabular}
\caption{Models and Sub-models active in the present study.}
\label{tab:activeModels}
\end{table}

\section{UV irradiance computation in CFD}

Let us consider a cylindrical UV lamp of radius $R_\lamp$ and length $L_\lamp$, centered at $z=0$ and its axis aligned along $z$ such that $z \in \left[-\frac{L_\lamp}{2}, \frac{L_\lamp}{2}\right]$ as it can bee seen from figure \ref{fig:BenchMarkDesign}.

Assuming that this UV lamp can emit diffusely and uniformly with total radiant exitance $M$ (\unit{\W\per\m\squared}), then the irradiance at a point $P$ in space can be computed using a \textit{view factors formulation approach} \cite{siegelhowell1992radiation, kowalski2009uvgi, modest2013radiative}.

\subsection{UV power irradiance formulation}

The irradiance at point $P$ is defined as $E_d$ (\unit{\W\per\m\squared}) such that:
\begin{equation}
E_d = \int_{A_\lamp} \frac{I(\mathbf{r}') \cos\theta \cos\theta'}{|\mathbf{r}-\mathbf{r}'|^2} \, dA',
\end{equation}
where:
\begin{itemize}
\item $\mathbf{r}$ is the observation point,
\item $\mathbf{r}'$ is a source point on the lamp surface,
\item $\theta$ is the angle between the surface normal at $P$ and incoming ray,
\item $\theta'$ is the emission angle at the lamp surface,
\item $A_\lamp$ is the lamp surface.

\end{itemize}

For a diffuse (\textit{Lambertian}) emitter, we have:
\begin{equation}
I = \frac{M}{\pi}
\end{equation}

Considering a cylindrical UV lamp of radius $R_\lamp$ and length $L_\lamp$, the UV irradiance is developed in the CFD solver by using a view factor $F_{dA \rightarrow A_\lamp}$ formulation \cite{siegelhowell1992radiation, kowalski2009uvgi, modest2013radiative} as the following:
\begin{equation}\label{eq:E_d}
E_d = M \, F_{p \rightarrow A_\lamp}
\end{equation}

The differential view factor from a differential receiver area $dA_i$ to the lamp surface is
\begin{equation}
dF_{p \rightarrow dA'} =
\frac{1}{\pi}
\frac{\cos\theta \cos\theta'}{r^2} \, dA',
\end{equation}
with $r = |\mathbf{r} - \mathbf{r}'|$.

The cylindrical UV lamp surface is decomposed into:
\begin{itemize}
\item Curved surface: $dA_c = R_\lamp \, d\phi \, dz$, and
\item End caps (optional): $dA_e = r \, dr \, d\phi$
\end{itemize}

For the lateral surface, one can write:
\begin{equation}
\mathbf{r}'(\phi,z) =
\begin{bmatrix}
R_\lamp\cos\phi \\
R_\lamp\sin\phi \\
z
\end{bmatrix},
\quad
z \in \left[-\frac{L_\lamp}{2}, \frac{L_\lamp}{2}\right]
\end{equation}

The final integral of the irradiance of the cylindrical UV lamp is computed as the following:

\begin{equation}
E_d =
\frac{M}{\pi}
\int_{-L_\lamp/2}^{L_\lamp/2}
\int_{0}^{2\pi}
\frac{\cos\theta \, \cos\theta'}{r^2}
R_\lamp \, d\phi \, dz
\end{equation}
where:
\begin{equation}
r = \left|\mathbf{r}_p - \mathbf{r}'(\phi,z)\right|
\end{equation}
and $\mathbf{r}_p = (x_p,y_p,z_p)$ is the observation point and,
\begin{equation}
r^2 =
(x_p - R_\lamp\cos\phi)^2 +
(y_p - R_\lamp\sin\phi)^2 +
(z_p - z)^2
\end{equation}

For a cylindrical surface with outward normal:
\begin{equation}
\mathbf{n}' =
\begin{bmatrix}
\cos\phi \\
\sin\phi \\
0
\end{bmatrix}
\end{equation}

Thus:
\begin{equation}
\cos\theta' =
\mathbf{n}' \cdot \frac{\mathbf{r}_p - \mathbf{r}'}{|\mathbf{r}_p - \mathbf{r}'|}
\end{equation}

For a planar receiver (surface normal $\mathbf{n}_p$):
\begin{equation}
\cos\theta =
\mathbf{n}_p \cdot \frac{\mathbf{r}' - \mathbf{r}_p}{|\mathbf{r}' - \mathbf{r}_p|}
\end{equation}

Then, the geometric view factor in eqn. \ref{eq:E_d} can be defined as:
\begin{equation}
F_{p \rightarrow \lamp} =
\frac{1}{\pi}
\int_{A_\lamp}
\frac{\cos\theta \cos\theta'}{r^2} \, dA'
\end{equation}

\section{Analysis of coronavirus diffusion inside airborne saliva droplets}

If $d_i(t)$ is the saliva droplet diameter that is time-dependent (e.g. due to different evaporation rates), then the Stokes--Einstein diffusion coefficient  $D_v(t)$ of a virus-in-saliva droplet can be estimated by:
\begin{equation}
D_v(t)~=~\frac {k_B T(t)} {3 \pi \mu(t) d_\capsid}
\end{equation}
where $d_\capsid$ is the virus capsid external mean diameter ($d_\capsid=\qty{100}{\nm}$), $k_B$ the Lattice Boltzmann constant, $\mu(t)$ the time-dependent liquid saliva viscosity and ${T_d}(t)$ is the air flow speed and temperature dependent effective temperature of the saliva droplet or particle.
As an example for \qty{100}{\um} saliva droplet in diameter, at $T=\qty{21}{\degreeCelsius}$ with dynamic viscosity as that of water ($\approx \qty{1}{\milli\Pa\s}$) and assuming still air, then a coronavirus of $d_\capsid=\qty{100}{\nm}$ will need about \qty{95}{\s} to diffuse from the center of the saliva droplet ($r=0$) towards its boundary ($r=R_{saliva}=\qty{50}{\um}$). 
Since Sapkota et al. \cite{Sapkota2026} proved experimentally that saliva droplets are subject to change in their morphology and in their effective viscosity due to water content loss by evaporation mechanisms and the remaining biological constituents. So, if one assumes a 3 times fold increase in the effective viscosity of a saliva droplet compared to water (e.g. \qty{3}{\milli\Pa\s}), and a twice reduction of the saliva droplet size, then this leads to a diffusion time of about $(95\times3)/2=\qty{125}{\s}$.
Chen et al. \cite{Chen2025} reported hydrated saliva viscosity up to \qty{100}{\milli\Pa\s} at low shear rates O(\qty{0.1}{\per\s}), but of course saliva effective viscosity can increase with evaporation becoming gel-like or solid droplets \cite{Sapkota2026} due to water loss and increase in protein residues and in mucin content. In other words, airborne dehydrated saliva droplets undergoing evaporation can witness an increase up to 10 to 100 times in viscosity compared to saliva droplets without evaporation, which is quasi-like impossible to measure in droplets flight mode.

As a summary of the above analysis, in hydrated saliva droplets of small sizes ($d<\qty{3}{\um}$), a \qty{100}{\nm} virion capsid can reach the droplets boundary within the order of magnitude of the total residence time of the droplet; e.g system of figure \ref{fig:BenchMarkDesign} at \qty{100}{\m\cubed\per\hour}. But in droplets of larger size (e.g. $d \geq \qty{3}{\um}$), diffusion starts to \textit{saturate} regardless of the droplet's size. For example, in dehydrated saliva of $d=\qty{20}{\um}$ and of $\mu=\qty{10}{\milli\Pa\s}$ (or a honey like viscosity at \qty{20}{\degreeCelsius}), a \qty{100}{\nm} virion capsid displacement will be limited to \qty{0.362}{\um} in \qty{0.5}{\s}. In other words, most biologically relevant droplets between 10 and \qty{500}{\um} are firmly in the diffusion-limited regime, not geometry-limited, over this timescale.

When it comes to non still saliva droplet moving in air, there will be a competition between advection and diffusion. But because, Sapkota et al. \cite{Sapkota2026} showed that saliva droplets under evaporation become very viscous as \textit{semi-solid} or \textit{gel-like} (see \cite{Sapkota2026}). Thus a very large increase in the effective viscosity, leads to the fact that advection will be very small compared to diffusion. In other words, this means that airflow will mainly moves the droplets externally, not internally mixing them, thus diffusion-dominated mechanisms with \textit{internally immobilized virions inside the saliva droplets}; e.g. no meaningful mixing will occur.

The above important conclusion of diffusion-dominated with \textit{internally immobilized virions inside the saliva droplets}, highlights the vital role of UV scattering in infected saliva droplets when exposed to UV irradiation field in an air flow dynamics system. Our primary conclusion in this context is that even if a saliva droplet is exposed to a high UV irradiation field, then UV light scattering energy will be importantly reduced in different local zones inside the droplet depending on its morphology and biological constituents. This what we will try to quantify and analyze by applying the DDA method and its implicit coupling to advanced CFD Euler-Lagrange platform.

\newpage

\begin{figure}[ht]
    \includegraphics[width=\textwidth]{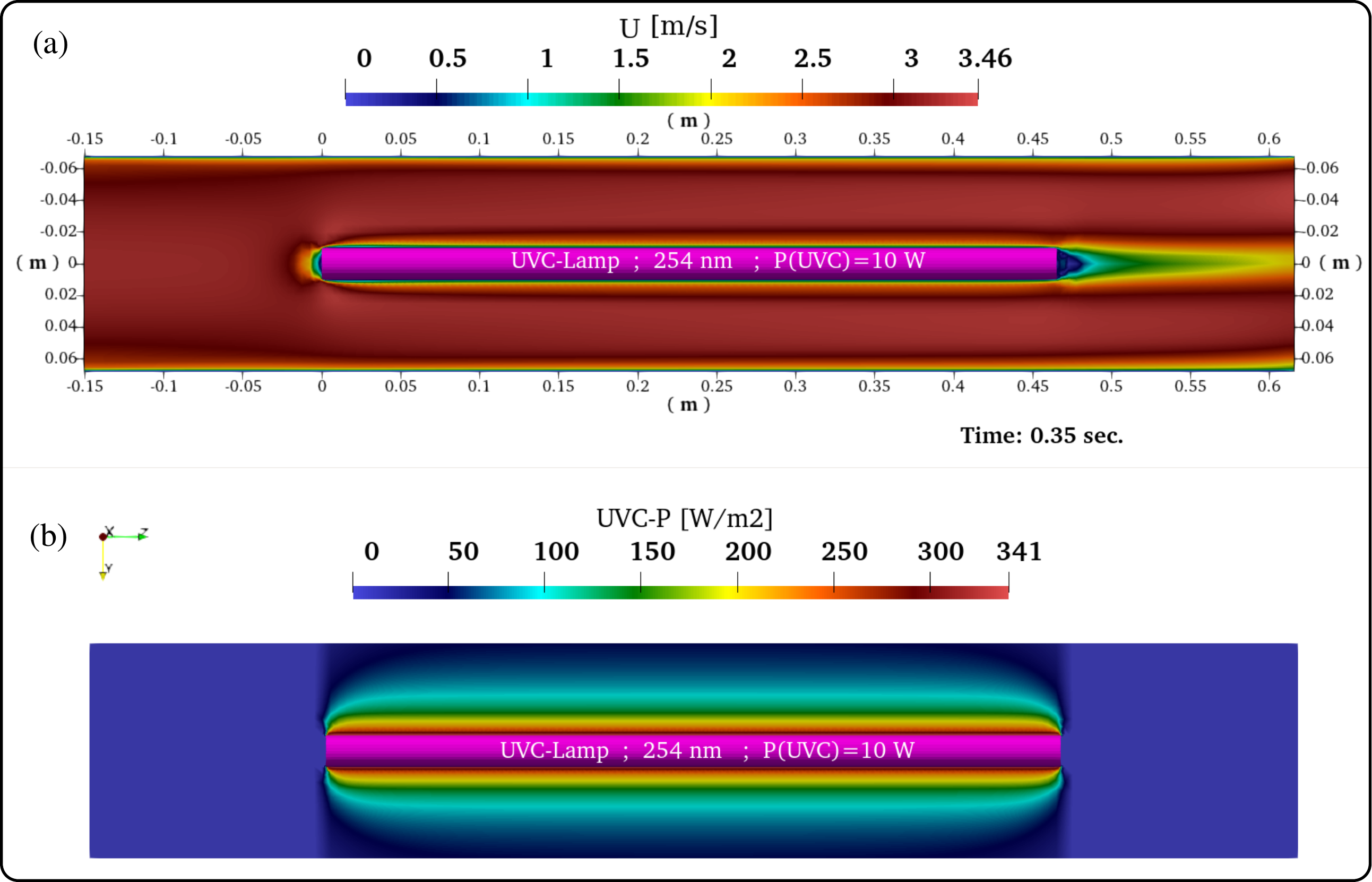}
        \caption{Coupled CFD of air flow with UV lamp irradiation for the predictions of infected saliva droplets dynamics in an air purifier benchmark design with a UV-C-Lamp of \qty{34.3}{\mW\per\cm\squared} corresponding to about 10 Watts of UV-C power. The UV lamp surface is considered to operate at \qty{38}{\degreeCelsius}. Results only showing: \textbf{(a)} Velocity field magnitude of the local air flow around the UV-C lamp at $t=\qty{0.35}{\s}$; \textbf{(b)} UV-C power field prediction within the UV-C air purifier for airborne virus inactivation by UV technology. Case study for air flow circulation at \qty{150}{\m\cubed\per\hour}, \qty{21}{\degreeCelsius} and 65\% relative humidity.}
    \label{fig:U_UVC_P}
\end{figure}

Figure \ref{fig:U_UVC_P} shows the CFD results of the velocity magnitude and local UV-C power flux density. The geometry represents the benchmark design/geometry as proposed in figure 3 of Alic et al. \cite{Alic2021ThreeStageInactivation} for UV air purifier;  see also figure \ref{fig:BenchMarkDesign} for the present authors generated schematic.

\section{UV-C light scattering in saliva: bulk slab versus single droplet at (\qty{254}{\nm})}

The early works of light scattering go back to 1871 and 1908 by Rayleigh \cite{Rayleigh1871SkyLight} and Mie \cite{Mie1908TrubeMedien}.
When it comes to models development in droplets with different properties, one can mention the huge developments made by Dombrovsky in early 2000's \cite{Dombrovsky2003SpectralDieselDroplets,Dombrovsky2003SemitransparentDroplet} applied in the fields of combustion of fuel droplets.

When it comes to saliva liquid, Bourgin et al. \cite{Bourgin2021SalivaUV} were the only ones so far to do measurements on human saliva interference with UV light. They reported UV absorbance spectra (\qtyrange{200}{700}{\nm}), UV attenuation at \qty{222}{\nm} and at \qty{254}{\nm}, dried saliva UV transmission, and average absorption coefficient at \qty{254}{\nm} $\approx$ \qty{6.7}{\per\cm}. Bourgin et al. \cite{Bourgin2021SalivaUV} did not separate scattering from absorption but provided an experimentally measured attenuation with their study limited to minimum scale of some millimeters saliva droplet on surfaces. Bourgin et al. \cite{Bourgin2021SalivaUV} concluded that a new artificial saliva recipe is needed for UV decontamination testing, which put into question all the previous measurements made before 2021. Sesti-Costa et al. \cite{SestiCosta2022UVSaliva} showed experimentally that saliva dramatically changes the UV dose required for viral inactivation, while Monika et al. \cite{Monika2025FarUVC} presented a comparative investigation of far-UV-C (\qty{222}{\nm}) versus germicidal UV-C (\qty{254}{\nm}) radiation against virus-laden aerosols of artificial human saliva. Lukose et al. \cite{Lukose2021Photonics} made an comprehensive review of optical properties and spectroscopy of saliva. 

Nevertheless, the huge late efforts between 2021 and 2025, no one yet investigated UV light scattering in airborne saliva droplets.

In the following, a focus is given to the order of magnitude analysis of UV light scattering  estimation in a spherical saliva droplet compared to a slab of saliva of small thickness when each is being exposed to a UV-C light irradiance.

\subsection{Single slab: An example of saliva slab \qty{2.85}{\mm} in thickness}

Let us assume a saliva slab thickness of \qty{2.85}{\mm} (same thickness as in measurements done by Ma et al. \cite{BenMa2021}), then if the whole saliva slab is irradiated with \qty{254}{\nm} UV-C light, one expects the following:
\begin{itemize}
\item The Primary attenuation mechanism is absorption (typically $>80 - 95\%$ of attenuation).
\item The Secondary mechanism is \textit{Mie scattering} from bacteria, cells, and debris \cite{Mie1908TrubeMedien,Bohren1983AbsorptionScattering}.
\item A Minor contribution will be as \textit{Rayleigh scattering} from proteins and dissolved molecules \cite{Rayleigh1871SkyLight}.
\end{itemize} 

A \qty{2.85}{\mm} saliva slab is thus optically thick at \qty{254}{\nm}. The transmitted irradiation intensity is expected to be very low because strong absorption by proteins and other biomolecules will be combined with scattering losses.

\subsection{Single spherical droplet: An example for a saliva droplet \qty{10}{\um} in diameter}

Now assuming a \qty{10}{\um} saliva droplet that is being exposed to UV-C irradiation, then the optical functioning is governed primarily by the \textit{Mie scattering} \cite{Mie1908TrubeMedien} and refraction of the droplet as a whole. Due to \qty{10}{\um} optical path, internal multiple scattering is thus negligible. Moreover, absorption is much less significant ($<10\%$) than in a \qty{2.85}{\mm} saliva slab. 

This means that accurate scattering by an entire micrometric saliva droplet must be solved by an advanced UV-C scattering technique. In that purpose the discrete dipole approximation (DDA) method \cite{Yurkin2007DDAReview,Yurkin2023Chapter} is adopted in this work and coupled to CFD Euler-Lagrange within a new advanced numerical 3D CFD-DDA platform.

The above scattering and absorption analysis leads also to primary conclusions as the following:
\begin{itemize}
    \item When modeling airborne virus inactivation, researchers should only rely on UV susceptibility values (\unit{\m\squared\per\J}) measured from UV disinfection experiments that are made on viruses inside bulk liquid slabs as a continuous medium exposed to UV irradiation. 
    \item Experimental measurements on the inactivation of airborne virus in dispersed media, like aerosols or airborne infected droplets in-air circulation systems, can be very misleading on the true mechanisms responsible for the inactivation of the airborne viruses; e.g. in saliva droplets. 
    \item In the coming sections, we will try to clearly discuss the mechanisms responsible for airborne viruses inactivation in air purifiers like systems. This is through advanced modeling and simulations within an innovative 3D CFD-DDA numerical platform.
\end{itemize}

\subsection{Droplet rotation in guided air flow: An example for a saliva droplet}

Saliva droplets when emitted by coughing or sneezing will undergo important rotations, but when guided within an air flow; e.g. by an air purifier system, can droplet's rotation persist?
In fact, the rotation can not persist indefinitely, because air viscosity exerts a torque that damps the rotation of the saliva droplet. For example, for $i$-th spherical saliva droplet of diameter $d_i$, applying Stokes' rotational drag on a sphere and Newton's equation for rotational motion, then the characteristic spin-down time $\tau_\spin$ can be approximated as the following:

\begin{equation}\label{eq:tauSpin}
    \tau_\spin = \frac{\rho_p {d_i}^2 }{60 \mu_air}
\end{equation}


Assuming $\rho_p \approx \qty{1000}{\kg\per\m\cubed}$ and $\mu_air \approx \qty{1.8e-5}{\Pa\s}$, then eqn. \eqref{eq:tauSpin} gives $\tau_\spin \approx \qty{0.09}{\ms}$. Similarly, for $d_i=\qty{100}{\um}$ we have $\tau_\spin \approx \qty{9}{\ms}$. 

In other words, assuming a 100 $\mu$ m droplet that is initially spinning with small $\omega_0$, and that enters at 1.91 m/s (100 $m^3/h$ in figure \ref{fig:BenchMarkDesign}) an axial air flow, then the droplet rotation or spinning will decay exponentially due to aerodynamic rotational drag following: $\omega(t)=\omega_0 e^{-t/\tau_{spin}}$. More practically, averaging over local shear rates for different positions within the diameter of the main cylindrical flow, e.g. if close or away from the walls, then the spinning will stop in about 5 relaxation times, thus $5 \times \tau_{spin}$, meaning in about 45 $ms$ ($=5 \times 9 ~ms$).
 
Based on the above analysis and for averaged $d_i=\qty{100}{\um}$, droplet's rotation can be assumed negligible in the present study, because $\tau_\spin/\tau_c << 1$ where $\tau_c$ is the mean characteristic inlet-to-outlet flow passage of the saliva droplets inside the UV air purifier system under investigation (see figure \ref{fig:BenchMarkDesign}), or the local residence time or period.

\section{Discrete dipole approximation}

Being applicable to objects of arbitrary inhomogeneous shape \cite{Yurkin2007DDAReview,Yurkin2023Chapter}, DDA is perfectly suitable for an infected saliva droplet. This method is based on volume discretization and the main computational part is solving for the internal electric field $\mathbf{E}(\mathbf{r})$. Any other measurable quantities can be obtained from it.

We are interested in the effect of droplet parameters (shape, size, orientation) on the UV dose obtained by virions. It is related to the absorbed power, which is determined by local field intensity and contrast of the refractive index. Assuming that the virion is a sphere with diameter much smaller than all other dimensions (the wavelength and droplet size) and refractive index $m_\capsid$, the absorbed power density (per virion volume) in vacuum \cite{Bohren1983AbsorptionScattering} is:
\begin{equation}\label{eq:pabs_vacuum}
  W_\mathrm{abs}^\mathrm{vac} = 
  \frac{\omega\varepsilon_0}{2}
  \frac{\pi d_\capsid^3}{6}
  \operatorname{Im}(m_\capsid^2)
  \left|\frac{3\mathbf{E}_0}{m_\capsid^2+1}\right|^2,
\end{equation}
where $\omega$ is the angular frequency of the UV light, $\varepsilon_0$ is the vacuum permittivity, and the expression in the square brackets represents the quasistatic limit of the field inside a sphere. The same reasoning works for a virion at position $\mathbf{r}$ inside a saliva droplet of refractive index $m$:
\begin{equation}\label{eq:pabs_droplet}
    W_\mathrm{abs}(\mathbf{r}) = 
    \frac{\omega\varepsilon_0}{2}
    \frac{\pi d_\capsid^3}{6}
    \operatorname{Im}(m_\capsid^2)
    \left|\frac{3m\mathbf{E}(\mathbf{r})}{m_\capsid^2+m^2}\right|^2,
\end{equation}
where the local electric field (external to the virion) is used. 

As explained in section \ref{sec:optmodel}, we use a fixed refractive index of $m=\complexnum{1.6+0.001i}$. Moreover, we assume $\operatorname{Re}{(m_\capsid})=\operatorname{Re}{(m)}$ (roughly corresponding to dried proteins), while specific value of $\operatorname{Im}{(m_\capsid})$ is not important while it is sufficiently smaller than the real part. Then the ratio of absorbed powers is given by:
\begin{equation}\label{eq:xi}
    \mathbf{\xi}(\mathbf{r}) \equiv 
    \frac{W_\mathrm{abs}(\mathbf{r})}{W_\mathrm{abs}^\mathrm{vac}} =
    \left[\frac{m^2+1}{3}\right]^2
    \frac{\left|\mathbf{E}(\mathbf{r})\right|^2}{\left|\mathbf{E}_0\right|^2},
\end{equation}
where the ratio of field intensities is available on the cubical grid of voxels employed in DDA and numerical prefactor equals 2.31.

DDA simulations are performed with the open-source code \texttt{ADDA} \cite{Yurkin2011ADDA}, version 1.5.0-alpha3. We used filtered-coupled-dipoles formulation of DDA \cite{Yurkin2010FCD} and Bi-CGStab(2) iterative solver with convergence threshold of $10^{-3}$. This combination is the most reliable for the largest considered size. We used a single particle shape, defined by STL file (see \ref{fig:irregularDropletGeometry} below). Depending on the desired size, the voxel grid is built using the point-in-polyhedron (\texttt{pip}) tool \cite{schuh2007pip} distributed with \texttt{ADDA}. We used the same grid for both wavelengths. It is based on the default voxel density in \texttt{ADDA} (10 voxels per wavelength inside the particle) for wavelength of \qty{254}{\nm}, but was slightly denser for \qty{280}{\nm}.

\section{Results and discussion}

\subsection{Dynamics of airborne saliva droplets}

\begin{figure}[ht]
    \includegraphics[width=\textwidth]{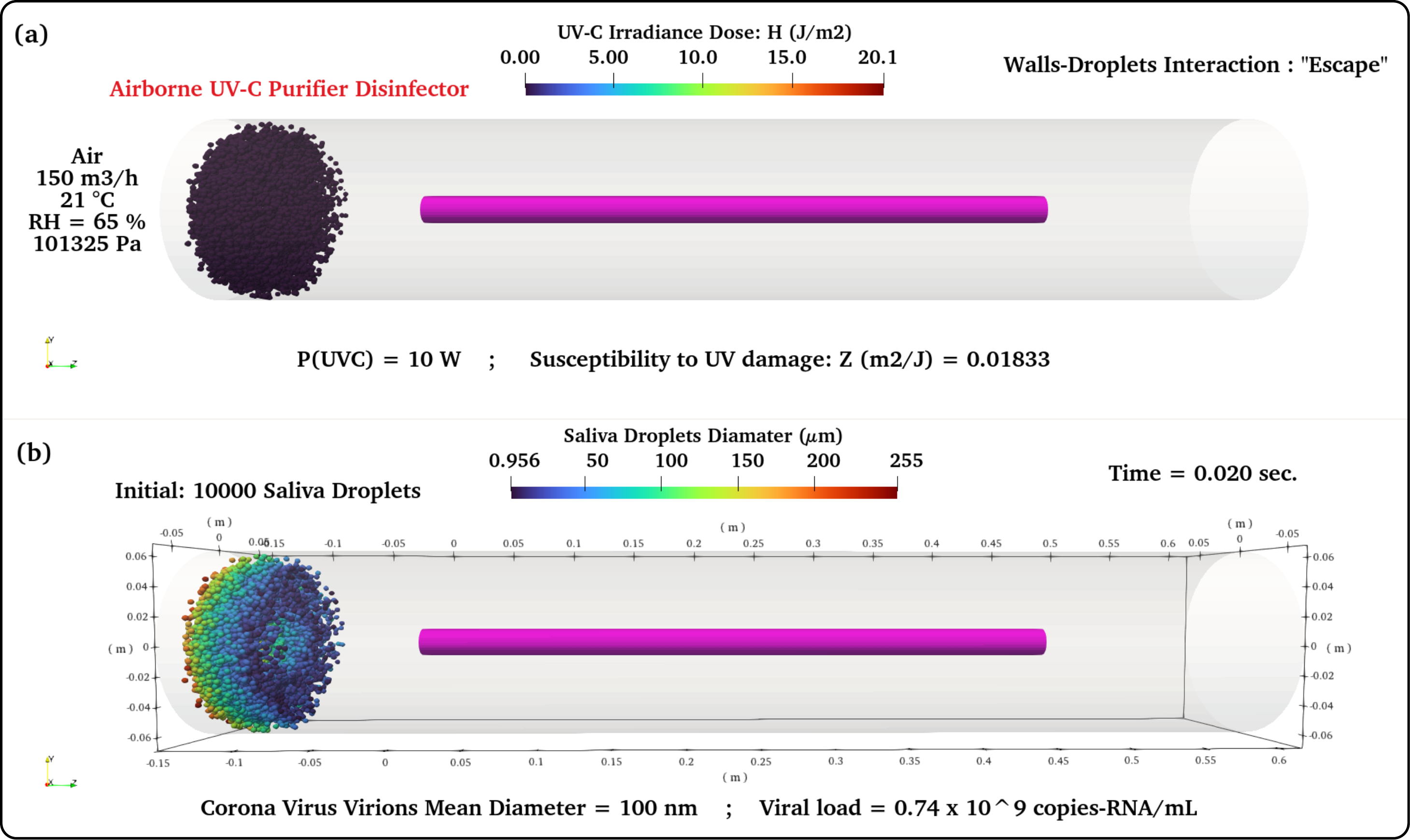}
    \caption{CFD predictions of infected saliva droplets dynamics and UV-C irradiance within an air purifier benchmark design with a UV-C-Lamp of \qty{34.3}{\mW\per\cm\squared}. Initial viral load of \num{0.74e9} copies-RNA/mL of Coronavirus \qty{100}{\nm} of capsid diameter corresponding to \num{3.7e-7} volume fraction in saliva droplets. The UV lamp surface is considered to operate at \qty{38}{\degreeCelsius}. Results at $\bf{t=\qty{0.02}{\s}}$ showing: \textbf{(a)} UV-C irradiance dose in \unit{\J\per\m\squared} accumulated within each droplet depending on its local trajectory history; \textbf{(b)} Droplets diameters local variation in \unit{\um} induced by the local evaporation rate of water content in saliva. Case study for $10^4$ initial saliva droplets with non-uniform diameters following the PDF distribution defined as PDF-a type shown in figure \ref{fig:PDF}. Case study for air flow circulation at \qty{150}{\m\cubed\per\hour}, \qty{21}{\degreeCelsius} and 65\% relative humidity. \href{https://drive.univ-rouen.fr/f/b528c4f3414441399425/}{download/play the multimedia video animation}.}
    \label{fig:CFD1}
\end{figure}

\begin{figure}[ht]
    \includegraphics[width=\textwidth]{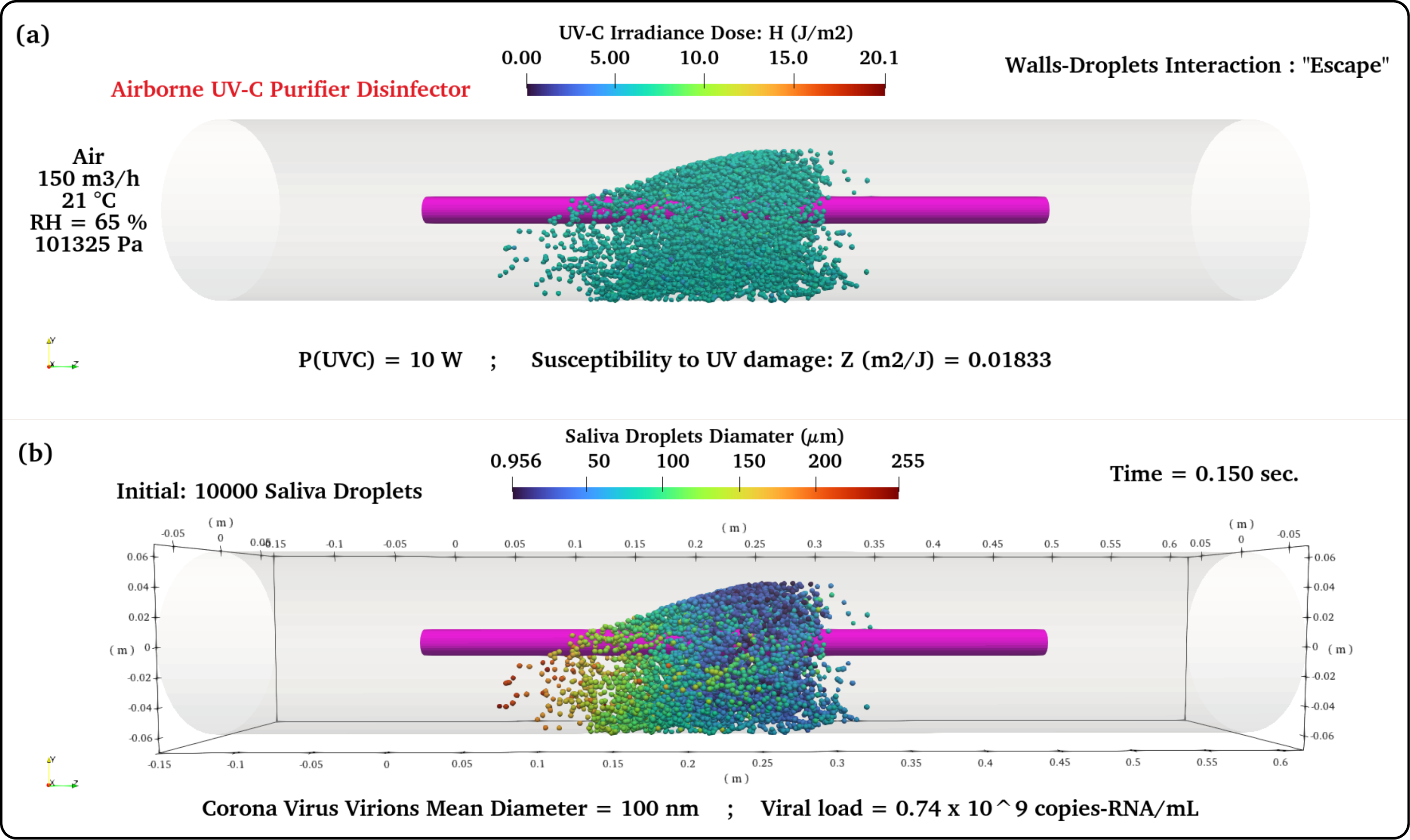}
    \caption{CFD predictions of infected saliva droplets dynamics and UV-C irradiance within an air purifier benchmark design. Same conditions as in figure \ref{fig:CFD1}. Results at $\bf{t=\qty{0.15}{\s}}$. \textbf{(a)} UV-C irradiance dose in \unit{\J\per\m\squared} accumulated within each droplet depending on its local trajectory history; \textbf{(b)} Droplets diameters local variation in \unit{\um} induced by the local evaporation rate of water content in saliva. \href{https://drive.univ-rouen.fr/f/b528c4f3414441399425/}{download/play the multimedia video animation}.}
    \label{fig:CFD2}
\end{figure}

\begin{figure}[ht]
    \includegraphics[width=\textwidth]{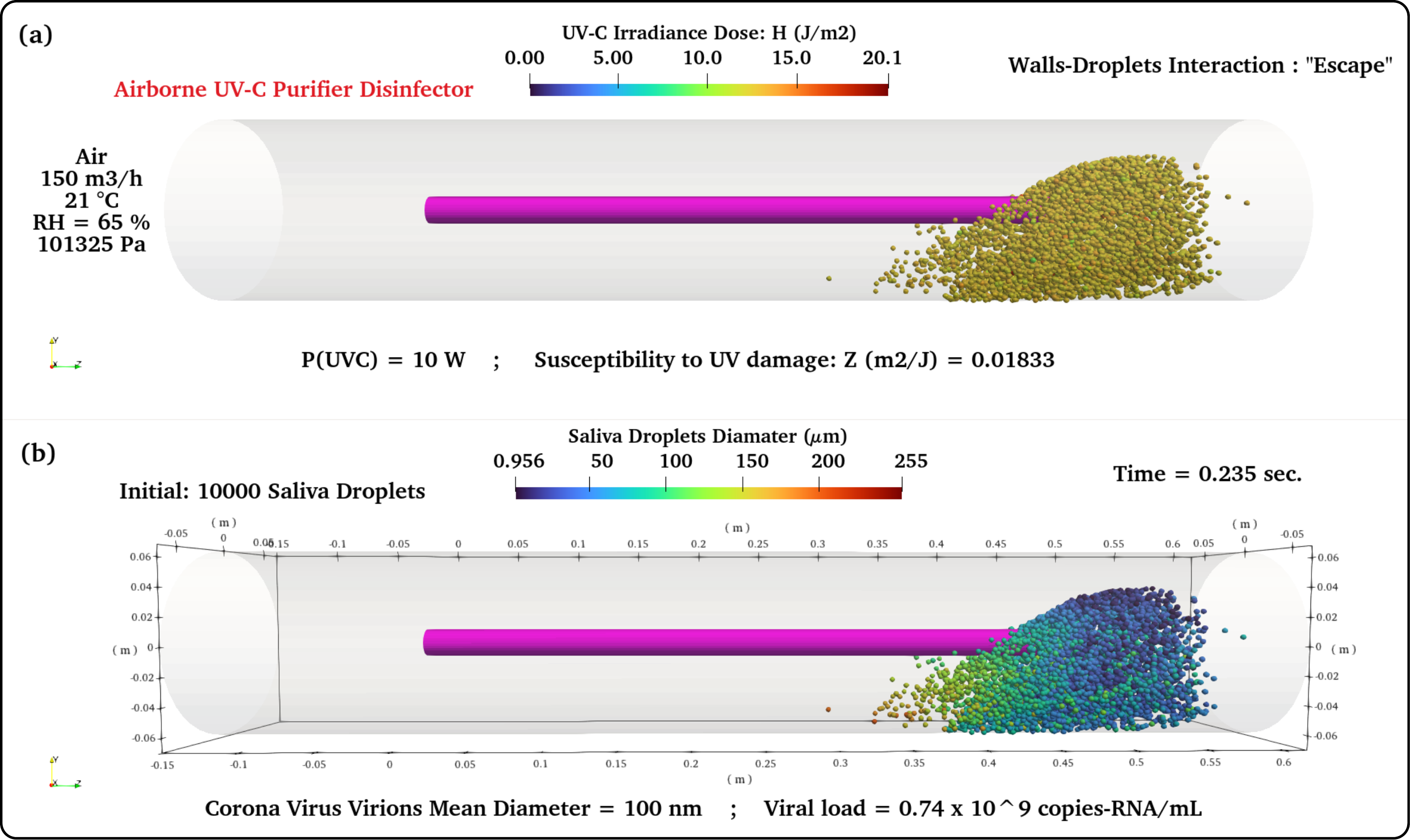}
    \caption{CFD predictions of infected saliva droplets dynamics and UV-C irradiance within an air purifier benchmark design. Same conditions as in figure \ref{fig:CFD1}. Results at $\bf{t=\qty{0.235}{\s}}$. \textbf{(a)} UV-C irradiance dose in \unit{\J\per\m\squared} accumulated within each droplet depending on its local trajectory history; \textbf{(b)} Droplets diameters local variation in \unit{\um} induced by the local evaporation rate of water content in saliva. \href{https://drive.univ-rouen.fr/f/b528c4f3414441399425/}{download/play the multimedia video animation}.}
    \label{fig:CFD3}
\end{figure}

The CFD predictions of infected saliva droplets dynamics and UV-C irradiance within an air purifier benchmark design with a UV-C-Lamp of \qty{34.3}{\mW\per\cm\squared} ar shown in figures \ref{fig:CFD1}, \ref{fig:CFD2} and \ref{fig:CFD3} (\href{https://drive.univ-rouen.fr/f/b528c4f3414441399425/}{download/play the multimedia video animation}.), at respective times of 0.02, 0.15 and \qty{0.235}{\s}. These results correspond to the case study of $10^4$ initial saliva droplets with non-uniform diameters following the distribution PDF-a presented in figure \ref{fig:PDF}. In this case, air flow circulation is at \qty{150}{\m\cubed\per\hour}, \qty{21}{\degreeCelsius} and 65\% relative humidity. In figure \ref{fig:CFD1}-a, the local UV-C irradiance dose in \unit{\J\per\m\squared} is accumulated within each droplet depending on its local trajectory history; while in figure \ref{fig:CFD1}-b, the droplets diameters local variations (in \unit{\um}) are induced by the local evaporation rate of water content in saliva particles. The initially $10^4$ infected saliva droplets contain a viral load of \num{7.4e8} copies-RNA/mL of Coronavirus \qty{100}{\nm} of capsid diameter corresponding to about \num{3.7e-7} volume fraction in the total number of airborne saliva droplets.

Figure \ref{fig:CFD1} clearly shows the entrance of the $10^3$ infected saliva droplets cloud at $t=\qty{0.02}{\s}$ for air flow rate of \qty{150}{\m\cubed\per\hour} and before being exposed locally to UV irradiance by the cylindrical UV-C lamp. This non exposure yet of droplets UV at $t=\qty{0.02}{\s}$ is clearly manifested by zero accumulated UV irradiance dose ($H=\qty{0}{\J\per\m\squared}$ in figure \ref{fig:CFD1}-a). Figure \ref{fig:CFD1}-b shows the local saliva droplets size distribution behavior at $t=\qty{0.02}{\s}$ after an initial distribution of type PDF-a shown in figure \ref{fig:PDF}-a. As expected, it can be seen from figure \ref{fig:CFD1}-b that leading saliva droplets within the flow direction from left to right are faster and lower in size, compared their trailing ones. After \qty{0.15}{\s}, figure \ref{fig:CFD2}-a shows how the saliva droplets cloud is now enveloping the UV lamp with more clear dispersion downwards due to gravity, and droplets have now a non-zero accumulated local UV dose. Figure \ref{fig:CFD2}-b illustrates how the saliva droplets size distribution at $t=\qty{0.15}{\s}$ is now developing due to local evaporation at the scale of each droplet. Similarly, figures \ref{fig:CFD3}-a and \ref{fig:CFD3}-b show how the fluid flow impacts the saliva droplets local dynamics and their local UV dose at $t=\qty{0.235}{\s}$ thus close to the residence time of the cloud inside the cylindrical air purifier/disinfection system.

\subsection{DDA simulations results}

\begin{figure}[ht]
    \includegraphics[width=\textwidth]{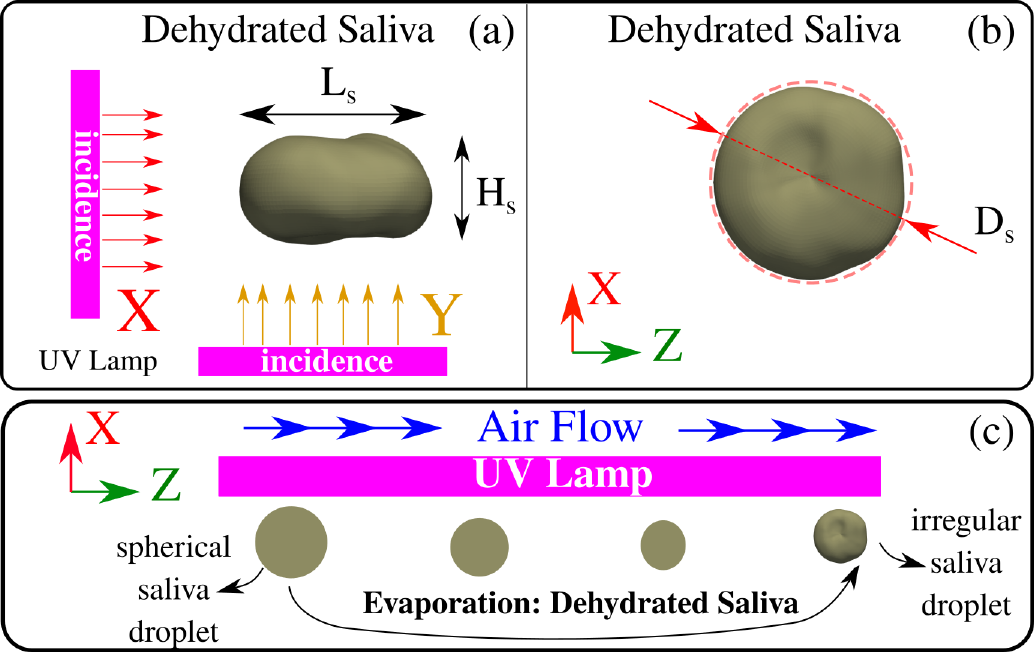}
    \caption{Dehydration of saliva droplet that is undergoing a shape change from regular sphere (undried or hydrated) to irregular shape (dried or dehydrated) droplet due to evaporation process in a fluid flow. 3D geometry as employed in the Discrete Dipole Approximation (DDA) solver to compute the UV light scattering 3D fields inside the droplet. (a) X,Y view; (b) X,Z view; (c) main air flow circulation direction. Note that five DDA simulations are considered with five different irregular droplet uniform scalings, such that $L_s \approx D_s \in \{1,2,3,4,5\}~\unit{\um}$ and $L_s/H_s=1.72$. $D_s$ is the diameter of a minimal sphere that envelops the irregular droplet; useful for comparisons with a spherical droplet of exact diameter $d\equiv D_s$. Two UV wavelengths are investigated: $\lambda=\qty{254}{\nm}$ and $\qty{280}{\nm}$; UV lamp power of 10 Watts corresponding to \qty{34.3}{\mW\per\cm\squared}.}
    \label{fig:irregularDropletGeometry}
\end{figure}

Figure \ref{fig:irregularDropletGeometry} shows the dehydration of saliva droplet that is undergoing a shape change from regular (sphere) to irregular shape dried droplet due to evaporation process in an air flow at \qty{100}{\m\cubed\per\hour}, \qty{21}{\degreeCelsius} and 65\% relative humidity. This 3D geometry was embedded in the Discrete Dipole Approximation (DDA) solver in order to compute the UV light scattering 3D fields inside the droplet. Droplet shape and orientation are illustrated in figures \ref{fig:irregularDropletGeometry}-a,  \ref{fig:irregularDropletGeometry}-b and \ref{fig:irregularDropletGeometry}-c depending on the air flow main direction. Five DDA different simulations were considered for five different sizes of irregular droplets with uniform X-Y-Z scaling, such that $L_s \approx D_s \in \{1,2,3,4,5\}~\unit{\um}$ and $L_s/H_s=1.72$. $D_s$ is the diameter of a minimal sphere that envelops the irregular saliva droplet; useful for comparisons with a spherical saliva droplet of exact diameter $d\equiv D_s$. Two UV wavelengths were investigated: $\lambda=\qty{254}{\nm}$ and $\qty{280}{\nm}$ with a UV lamp power of \qty{10}{\W} corresponding to \qty{34.3}{\mW\per\cm\squared}.

\begin{figure}[ht]
    \includegraphics[width=\textwidth]{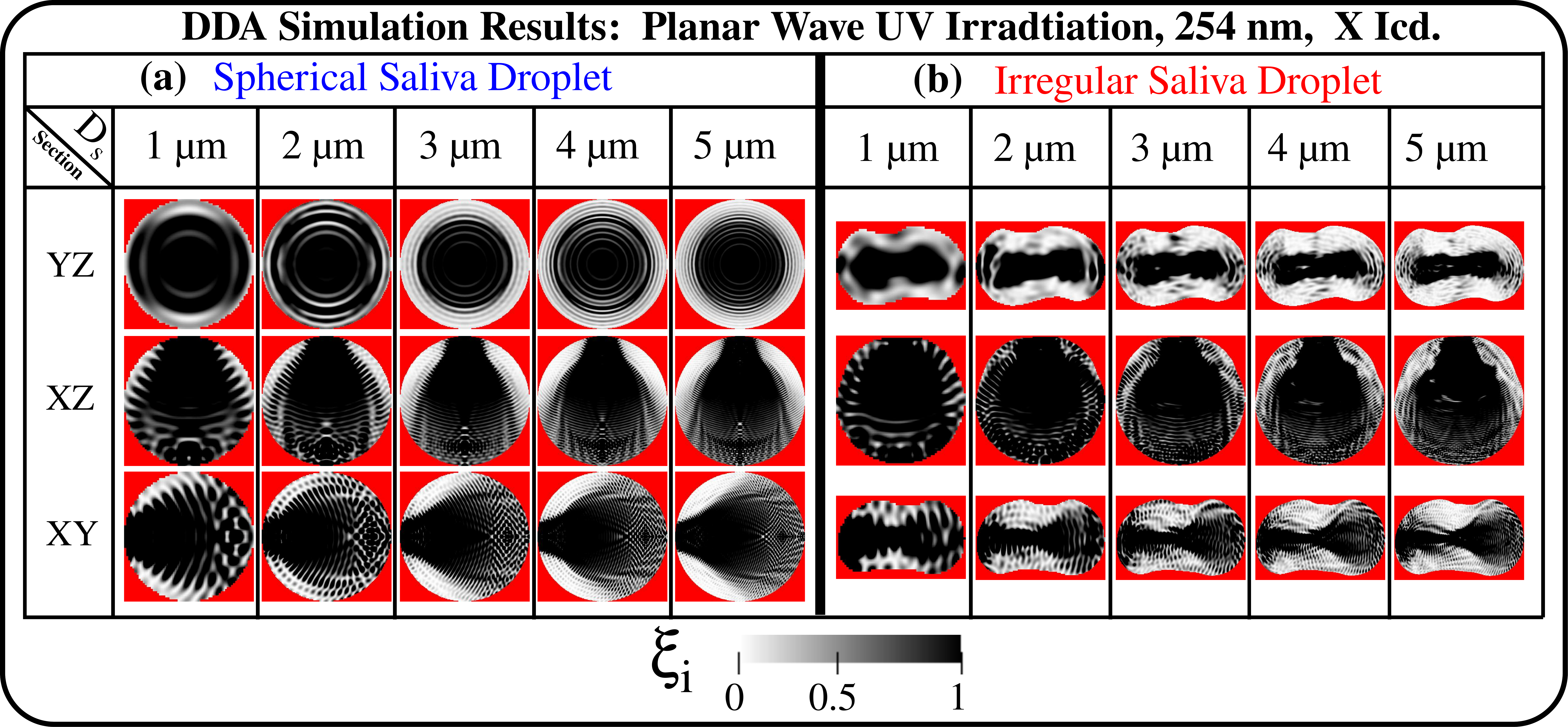}
    \caption{Discrete Dipole Approximation (DDA) results for UV scattering in both spherical and irregular saliva droplets of effective sizes $D_s \in [1,5]~\unit{\um}$. Results for UV light at $\lambda=\qty{254}{\nm}$. An example of the DDA simulation results for UV planar wave incidence along X-axis with refractive index $m=\complexnum{1.6+0.001i}$. (a) Spherical saliva droplet shape; (b) Irregular saliva droplet shape that can be induced by evaporation and local flow dynamics; see figure \ref{fig:irregularDropletGeometry}.}
    \label{fig:sphere_drop_size1_X_254nm}
\end{figure}

\begin{figure}[ht]
    \includegraphics[width=\textwidth]{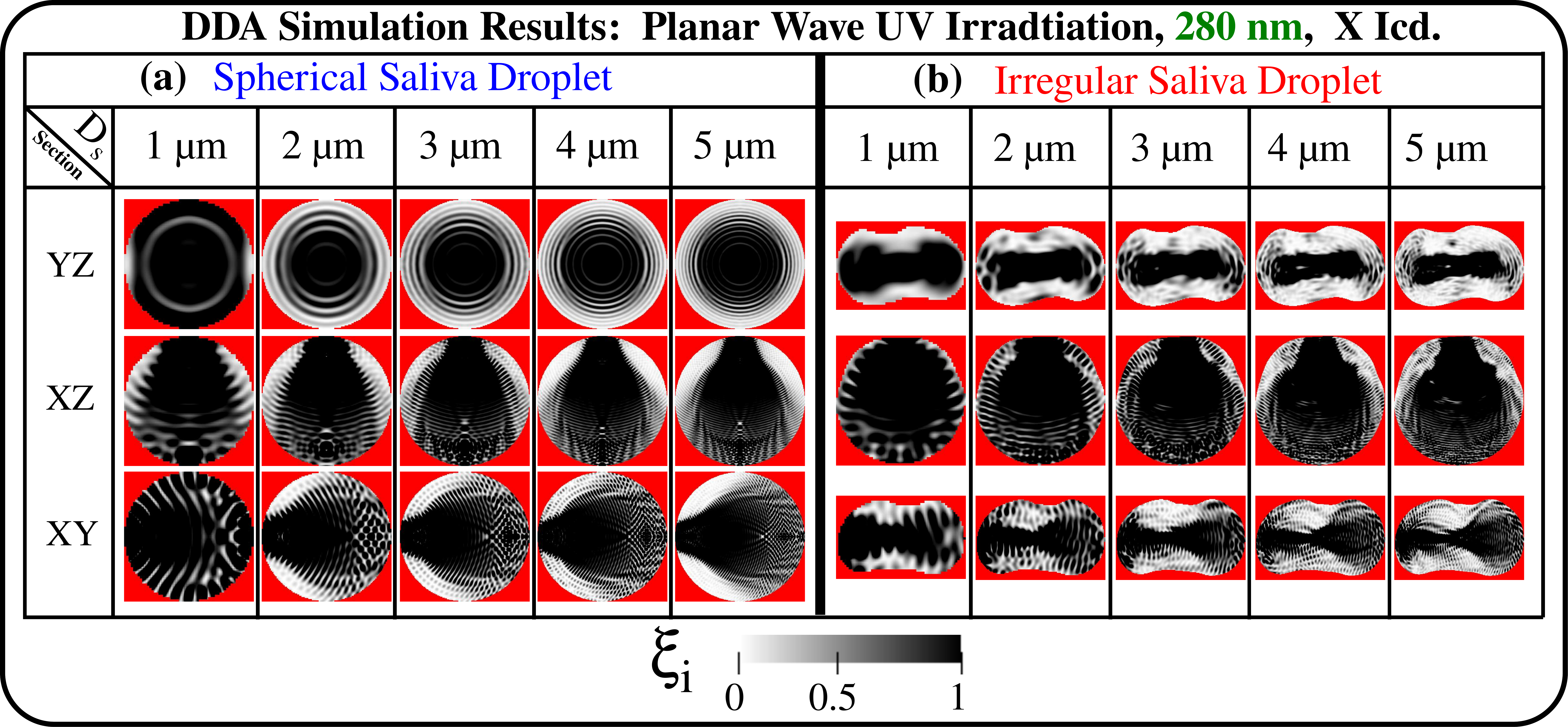}
    \caption{Discrete Dipole Approximation (DDA) results for UV scattering in both spherical and irregular saliva droplets of effective sizes $D_s \in [1,5]~\unit{\um}$. Results for UV light at $\lambda=\qty{280}{\nm}$. An example of the DDA simulation results for UV planar wave incidence along X-axis with refractive index $m=\complexnum{1.6+0.001i}$. (a) Spherical saliva droplet shape; (b) Irregular saliva droplet shape that can be induced by evaporation and local flow dynamics; see figure \ref{fig:irregularDropletGeometry}.}
    \label{fig:sphere_drop_size1_X_280nm}
\end{figure}

Figures \ref{fig:sphere_drop_size1_X_254nm} and \ref{fig:sphere_drop_size1_X_280nm} show an example of the DDA simulation results at two respective UV wavelengths $\lambda=\qty{254}{\nm}$ and $\qty{280}{\nm}$ and for droplets with characteristic sizes $D_s\in [1,5]~\unit{\um}$. Figure \ref{fig:sphere_drop_size1_X_254nm}-a correspond to DDA results for a spherical airborne saliva droplets while Figures \ref{fig:sphere_drop_size1_X_254nm}-b corresponds to DDA results for an irregular shaped airborne saliva droplets that can be induced by the evaporation process of water content or flow dynamics conditions. It can be clearly observed that a non-homogeneous UV scattering occurs with dimensionless UV energy field intensities $\xi$ that are lower than unity in local zones. Value of $\xi<1$ in figures \ref{fig:sphere_drop_size1_X_254nm} and \ref{fig:sphere_drop_size1_X_280nm} mean that the incidence UV irradiance power will be in fact reduced in many local zones inside the airborne saliva droplets due to UV light scattering. Moreover, the local spacing at low $\xi$ bands can be as large as \qty{0.5}{\um} in large-sized saliva droplets; e.g. $D_s \geq \qty{5}{\um}$.
For hydrated saliva, Chen et al. \cite{Chen2025} reported saliva viscosity up to \qty{100}{\milli\Pa\s} at low shear rates O(\qty{0.1}{\per\s}). But saliva droplet's effective viscosity can importantly increase with the evaporation process; e.g. saliva becomes gel-like or solid \cite{Sapkota2026} due to water content loss and an increase in the mucin and protein residues. Thus, dehydrated saliva droplet can witness an increase up to 10 to 100 times in viscosity compared to saliva droplet without evaporation.
Let us assume a mean saliva droplet of \qty{20}{\um} in diameter and \qty{10}{\Pa\s} in viscosity (e.g. honey at \qty{20}{\degreeCelsius}), this means that the displacement of a \qty{100}{\nm} virus capsid per droplet's total residence time of \qty{0.3}{\s} is about \qty{0.15}{\um}. Figures \ref{fig:sphere_drop_size1_X_254nm}-b and \ref{fig:sphere_drop_size1_X_280nm}-b clearly show that UV scattering phenomenon with $\xi<1$ can exist in numerous band zones (white color) of thickness much larger than \qty{0.15}{\um} in large sized droplets. In other words, these white-color bands represent indeed true high risk zones for virus survival that require a correction in the effective UV dose $H_i$ received by an i-th droplet; for example a correction in the log reduction law for the inactivation of viruses as we will show in the following section \ref{sec:uv_laplace}. 

\subsection{Virion diffusion in dehydrated saliva}

\begin{figure}[ht]
\includegraphics[width=0.7\textwidth]{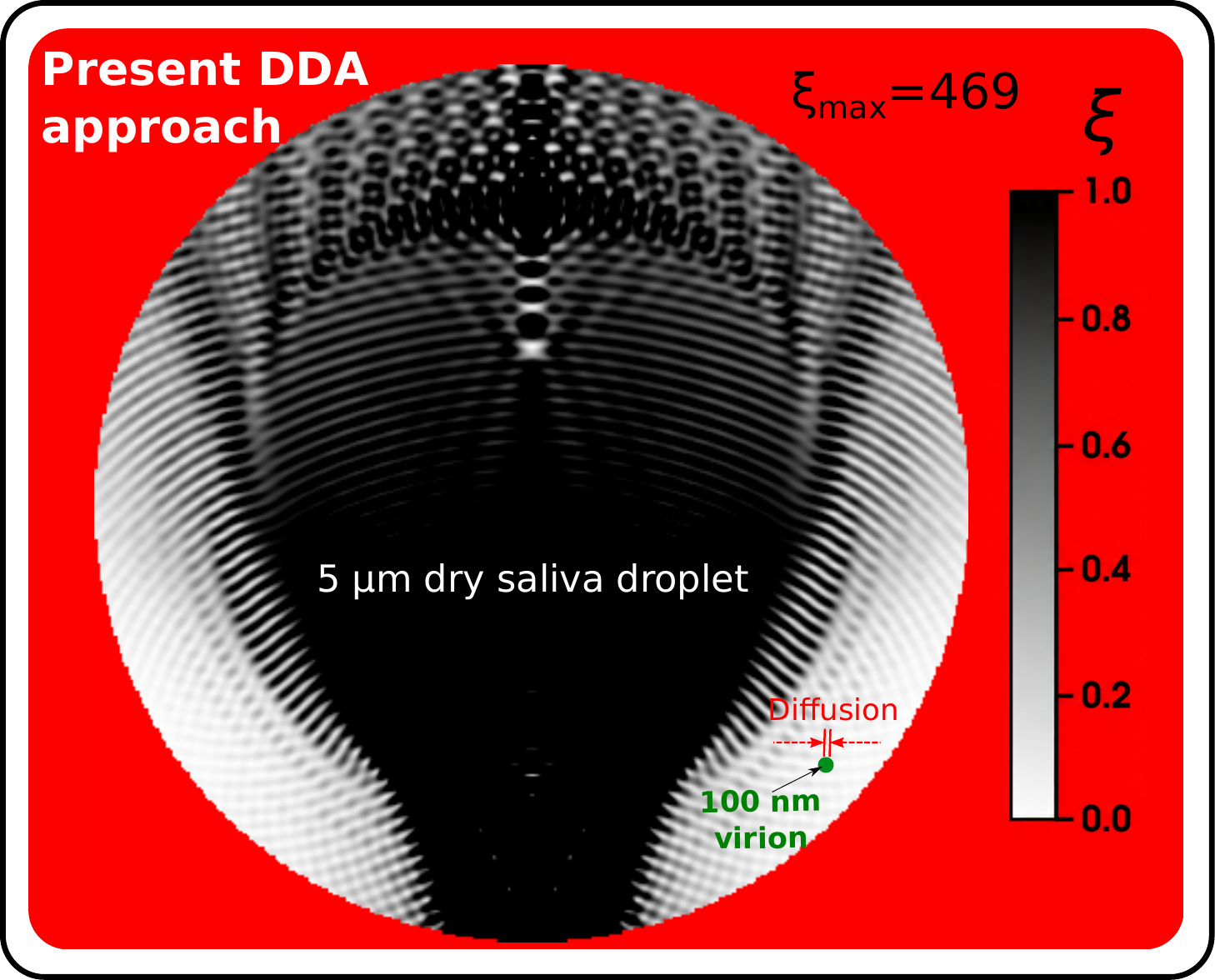}
\caption{Limited virus diffusion in a in a dry spherical saliva droplet with UV scattering given by DDA simulations (\num{15.2e+6} data points) for normalized intensity $\xi(\mathbf{r})$ of UV-C scattering field, see eqn. \ref{eq:xi}, in \qty{5}{\um} dry saliva droplet irradiated at $\lambda=\qty{254}{\nm}$ by a UV planar wave incidence. Negligible diffusion compared to the virion capsid size and the residence time of \qty{5}{\um} the spherical saliva droplet flying time history; \href{https://drive.univ-rouen.fr/f/e41019825f2146ed97fe/}{See download/play video animation}.}
    \label{fig:Diffusion_of_virion_in_dehydrated_saliva}
\end{figure}

Figure \ref{fig:Diffusion_of_virion_in_dehydrated_saliva} shows an example of limited Brownian diffusion of \qty{100}{\nm} virion capsid, the point close to the arrow, inside a dry saliva droplet \qty{5}{\um} in diameter. The white-black back ground image projects the DDA results for UV scattering at $\lambda=\qty{254}{\nm}$ for UV planar wave incidence along X-axis with refractive index $m=\complexnum{1.6+0.001i}$ (ZX cross-section of figure \ref{fig:sphere_drop_size1_X_254nm}, $D_s=\qty{5}{\um}$). Due to light scattering and limited diffusion, this virion stays approximately in its initial white-colored zone; thus it stays in a local zone of reduced incidence UV irradiance power represented by $\xi < 1$. \href{https://drive.univ-rouen.fr/f/e41019825f2146ed97fe/}{Download/play the multimedia video animation}.

\section{Virus Inactivation Law in a Spatially Non-Uniform UV Radiation Field}
\label{sec:uv_laplace}

\subsection{First-Order UV Inactivation Kinetics}

The inactivation of viruses by ultraviolet (UV) radiation can be commonly described using a first-order kinetic model. In its simplest form, the total fraction of infectious virus copies remaining (or survived) after exposure to a UV dose $H$ can be defined by the Chick--Watson \cite{kowalski2009uvgi,Walker2007} law:

\begin{equation}
	\frac{N_s(t)}{N_0} = e^{-Z H(t)}
	\label{eq:classical_inactivation}
\end{equation}

where $N_0$ is the total initial concentration of infectious viruses, $N_s$ is the total virus concentration survived or remaining after UV exposure and $Z$ is the UV susceptibility coefficient of the virus. The susceptibility coefficient has units of inverse dose, for example $\mathrm{m^2\,J^{-1}}$ when the UV dose is expressed in $\mathrm{J\,m^{-2}}$. 

$H$ is the time-dependent UV dose defined as:
\begin{equation}
H(t) = \int_{0}^{t} E_p(x,y,z) ~dt,
\end{equation}
where $E_p$ is the local UV irradiance in \unit{\W\per\m\squared} and $t$ is the exposure time in \unit{\s}.

Equation~\ref{eq:classical_inactivation} implicitly \textbf{\textit{assumes}} that a UV dose is spatially uniform throughout the host droplet, or the volume containing the viruses. This assumption is not valid for a saliva droplet subject to UV scattering that will produce a spatially heterogeneous UV intensity field, resulting in different local UV doses experienced by the viruses at different local positions within a saliva droplet.

\subsection{The susceptibility constant \texorpdfstring{$Z$}{Z}}

In fact, the susceptibility constant $Z$ depends on several factors such as: the UV wavelength, virus species, surrounding medium (air, water, surface), humidity and the environmental conditions.
A larger value of $Z$ indicates greater sensitivity to UV radiation, requiring a lower dose to inactivate the same number of virus copies.

The literature showed in very rare studies that viruses are more susceptible to UV irradiation when suspended in airborne droplets  compared to when suspended in a bulk liquid. But this indeed depends on the experimental setup because the reported rare values of $Z$ are very rare and disperse; only four experiments with early one goes back to Jensen 1964 \cite{Jensen1964} followed by three others \cite{Walker2007,Bedell2016,Welch2018}. 

For the above reason, we highlight that for engineering designs applications, engineering should reply more on $Z$ values measured in bulk liquid samples \cite{SestiCosta2022UVSaliva,Kariwa2004,DARNELL2004}, compared to $Z$ values measured in airborne experiments \cite{Jensen1964,Walker2007,Bedell2016,Welch2018}. This is because experiments done in bulk liquid samples are very much less independent of the experimental bench design. Usually bench setups are very much more complex in airborne and aerosols experiments involving virus inactivation, compared to those conducted in bulk liquid samples.

In the present research, two scenarios are investigated in the 3D CFD-DDA platform: $Z=\qty{0.0183}{\m\squared\per\J}$ and $Z=\qty{0.377}{\m\squared\per\J}$ as reported respectively in bulk liquid and in airborne aerosol experiments (see Beggs et al. \cite{Beggs2020}). In other words, this chosen lower and upper bounds values of $Z$ allow investigating both the best- and the worst-case scenarios for disinfection.

\subsection{UV scattering in airborne infected droplets}

\citet{Doughty2021} were the firsts to derive relative local UV intensities (denoted $R_i$ for planar UV waves at $\lambda=260~nm$ and $\lambda=302~nm$) in dried virion infected droplets. They conducted modeling and simulations using the Multi-Sphere T-Matrix (MSTM) method that provided solutions to Maxwell’s equations for a case included 12000 non-overlapping spherical virions (100 nm in diameter) inside a spherical droplet (host). Their results, see figure-2b in \cite{Doughty2021}, are in good agreement and very similar to our local normalized intensity field $\xi(\mathbf{r})$ shown in figure \ref{fig:Diffusion_of_virion_in_dehydrated_saliva} that is obtained by DDA simulations \cite{Yurkin2011ADDA,Yurkin2007DDAReview}. DDA methods allows millions of data points inside a droplet, but the MSTM is limited only to thousands of virions inside a droplet.

The similarity between the present $\xi_i(\mathbf{r})$ field and the $R_i$ field computed by \citet{Doughty2021}, allows us to assume that a $\xi_i(\mathbf{r})$ value corresponds to a UV local intensity value received by an i-th virion located a specific position $\mathbf{r=r_{virion}}$ inside the droplet. 

It is also good to highlight that \citet{Doughty2021} assumed a random distribution of virions inside the droplet and they did not couple their fluence $F$ or UV-dose and their relative local UV intensities $R_i$ with CFD or with local fluid flow dynamics.

In our works, the local normalized intensity $\xi(\mathbf{r})$, see equation \ref{eq:xi}, is computed through the DDA simulations, keeping in mind that $\xi>1$ correspond to local intensity enhancement, whereas $\xi<1$ correspond to local intensity attenuation relative to the incident field.

If the local UV dose is proportional to the local intensity, then the dose experienced at a position $\mathbf{r}$ inside an infected i-th saliva droplet can be coupled with local fluid flow dynamics as:

\begin{equation}
	H_i(\mathbf{r},t) = H_i(t) \xi_i(\mathbf{r})
	\label{eq:local_dose}
\end{equation}

where $H_i(t)$ represents the fluid flow-dynamics-dependent local UV dose in i-th droplet corresponding to the incident field in the absence of the spatial field modulation considered here.

Following the works by ~\citet{Doughty2021} and substituting eqn.~\ref{eq:local_dose} into the first-order inactivation law in eqn. \ref{eq:classical_inactivation}, we can thus define a space-time-dependent local surviving number $\left[N_s(\mathbf{r,t})\right]_i$ of virions in an infected i-th saliva droplet as the following:

\begin{equation}
	\left[\frac{N(\mathbf{r},t)}{N_0(\mathbf{r})}\right]_i
	=
	e^{-Z H_i(t) \xi_i(\mathbf{r})}
	\label{eq:local_survival}
\end{equation}

Eqn.~\ref{eq:local_survival} clearly demonstrates that the spatially varying electromagnetic field directly translates into a spatially varying viral inactivation probability that couples both DDA and CFD simulations effects for both local fluid flow dynamics and local UV intensity inside each virus-infected saliva droplet.

\subsection{Local UV Intensity Distribution functions $p(\xi)$ from DDA simulations}

\begin{figure}[ht]
    \includegraphics[width=\textwidth]{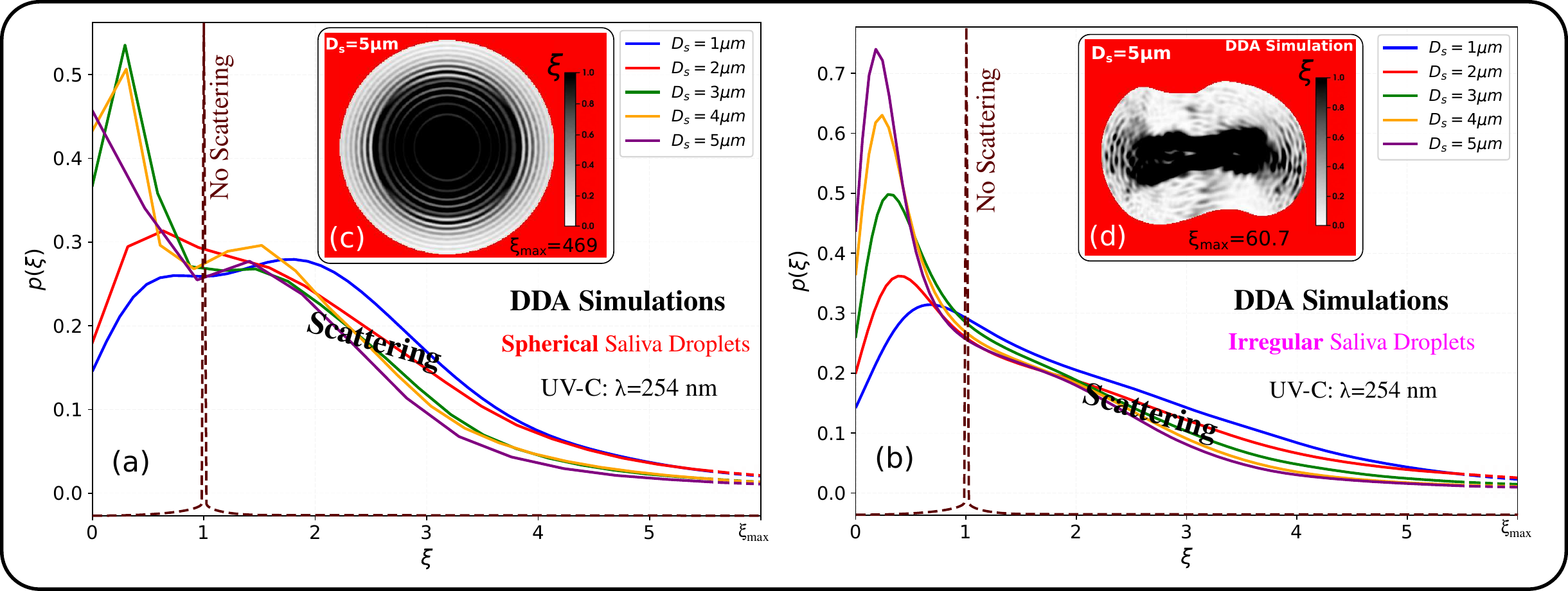}
    \caption{Discrete Dipole Approximation (DDA) results for UV scattering in spherical (a,c) and irregular shaped (b,d) saliva droplets of effective sizes $D_s \in [1,5]~\unit{\um}$. Results of the distribution function $p(\xi)$ as function of the normalized UV intensity field $\xi$ inside different sizes/shapes of airborne saliva droplets. UV exposure at X-incidence planar wave with refractive index $m=\complexnum{1.6+0.001i}$ and $\lambda=\qty{254}{\nm}$. (a) Spherical saliva droplets with an example in (c) for a YZ-cross section at $D_s=5 \mu m$; (b) Irregular shaped saliva droplets with an example in (d) for a YZ-cross section at $D_s=5 \mu m$; see figure \ref{fig:irregularDropletGeometry}.}
    \label{fig:DDA_p_of_xi_dried_vs_undried_saliva}
\end{figure}

Figure \ref{fig:DDA_p_of_xi_dried_vs_undried_saliva} shows an example of the local UV intensity distribution functions $p(\xi)$ as obtained from DDA simulations for both spherical saliva droplets, see figure \ref{fig:DDA_p_of_xi_dried_vs_undried_saliva}-c, and irregular shaped saliva droplets, see figure \ref{fig:DDA_p_of_xi_dried_vs_undried_saliva}-d. Note that a change in the shape of saliva, e.g. due to evaporation, can affect the local UV intensity distribution function $p(\xi)$ as it can be clearly observed from figures \ref{fig:DDA_p_of_xi_dried_vs_undried_saliva}-a and \ref{fig:DDA_p_of_xi_dried_vs_undried_saliva}-b. Irregular shaped saliva droplets can witness higher $p(\xi)$ peaks at low $\xi$ values as it can be seen from the peaks of curves in figure \ref{fig:DDA_p_of_xi_dried_vs_undried_saliva}-b that shift away from $\xi=1$ to lower values of $\xi$ with increasing droplet's size. This means a higher probability of virus copies survival in irregular shaped dried airborne saliva droplets. This also highlights the importance of evaporation and UV scattering in airborne saliva droplets dynamics to be both coupled and considered; e.g. by employing CFD-based UV irradiance investigations.

\subsection{Volume-Averaged Survival Fraction}

Knowing that viruses are distributed throughout an i-th saliva droplet, then the experimentally observed surviving fraction corresponds to an average over the droplet volume. Assuming a uniform initial virus concentration, the surviving fraction $\left[N_s/N_0\right]_i$ can thus be written as:

\begin{equation}
	\left[\frac{N_s(\mathbf{r},t)}{N_0(\mathbf{r})}\right]_i
	=
	\frac{1}{V}
	\int_V
	e^{-Z H_i(t) \xi_i(\mathbf{r})}
	\,dV
	\label{eq:volume_average}
\end{equation}

Moreover, if the initial virus concentration in a droplet is spatially non-uniform, $N_0(\mathbf{r})$ can be included as a weighting function such that:

\begin{equation}
	\left[\frac{N_s}{N_0}\right]_i
	=
	\frac{
		\displaystyle
		\int_V
		N_0(\mathbf{r})
		e^{-Z H_i(t) \xi_i(\mathbf{r})}
		\,dV
	}{
		\displaystyle
		\int_V N_0(\mathbf{r})\,dV
	}
	\label{eq:weighted_volume_average}
\end{equation}

Note that, for a uniformly distributed viral population, eqn. (\ref{eq:weighted_volume_average}) reduces to eqn. (\ref{eq:volume_average}).

\subsection{Probability Distribution of the Local Field Enhancement}

The spatially varying quantity $\xi(\mathbf{r})$ can alternatively be represented by a probability density function (PDF), denoted by $p_\xi(\xi)$. This PDF describes the fraction of the droplet volume experiencing a particular value of the normalized local intensity. This PDF can be defined as:

\begin{equation}
	p_\xi(\xi)
	=
	\frac{1}{V}
	\int_V
	\delta
	\left[
	\xi-\xi(\mathbf{r})
	\right]
	dV,
	\label{eq:pdf_xi}
\end{equation}

where $\delta(\cdot)$ is the Dirac delta function. The above leads to:

\begin{equation}
	\int_{0}^{\infty} p_\xi(\xi)\,d\xi = 1
	\label{eq:pdf_normalization}
\end{equation}

Using the PDF defined in eqn. (\ref{eq:pdf_xi}), the volume integral in eqn. (\ref{eq:volume_average}) can be transformed into an integral over the local values of $\xi$ in an i-th droplet such that:

\begin{equation}
	\left[\frac{N_s(\mathbf{r},t)}{N_0}\right]_i
	=
	\int_0^\infty
	e^{-Z H_i(t) \xi_i}
	[p_\xi(\xi)]_i
	\,d\xi
	\label{eq:survival_pdf}
\end{equation}

This new formulation in eqn. (\ref{eq:survival_pdf}) clearly separates now the problem into 2 major coupled physics components:
\begin{itemize}
	\item  The viral inactivation kinetics in an i-th droplet, carrier-flow-dependent, which are described by the term $e^{-Z H_i(t) \xi_i}$, and
    \item The electromagnetic scattering properties of an i-th saliva droplet which are contained in the probability distribution $[p_\xi(\xi)]_i$.
\end{itemize}

\subsection{Laplace-Transform Representation}

Eqn. (\ref{eq:survival_pdf}) has also a mathematical form of a Laplace transform of a function $p_\xi(\xi)$ defined as:

\begin{equation}
	\mathcal{L}\{p_\xi\}(s)
	=
	\int_0^\infty
	e^{-s\xi}
	p_\xi(\xi)
	\,d\xi
	\label{eq:laplace_definition}
\end{equation}

Thus introducing a dimensionless Laplace-transform argument:

\begin{equation}
	s = Z H(t),
	\label{eq:laplace_variable}
\end{equation}

Then, eqn. (\ref{eq:survival_pdf}) can be rewritten in a more compact form as:

\begin{equation}
		\left[\frac{N_s}{N_0}\right]_i
		=
		\mathcal{L}
		\left\{
		[p_\xi(\xi)]_i
		\right\}
		(Z H_i(t))
	\label{eq:laplace_survival}
\end{equation}

Thus, the overall viral survival fraction in a droplet is the Laplace transform of the probability distribution of the local UV intensity enhancement factor $\xi$, evaluated at $s=Z H(t)$.

Eqn. (\ref{eq:laplace_survival}) can thus be applied to compute $\left[\frac{N_s}{N_0}\right]_i$ which is the time-dependent virus survival fraction in each infected i-th saliva droplet.

The total virus copies survival fraction $\frac{N_s}{N_0}$ can be computed at time $t$ from the airborne saliva droplets cloud dynamics in the CFD code such that:

\begin{equation}
    \frac{N_s}{N_0} = \sum_{i=1}^{n(t)} \left[\frac{N_s}{N_0}\right]_i
\end{equation}

The above derivation approach provides a direct mathematical connection between the local UV field calculated using DDA and the macroscopic viral inactivation response, all taking into account the local fluid flow behavior resolved by the CFD code.

In the coming sections, we will clearly show how UV optical scattering obtained from DDa simulations can indeed produce within CFD coupling \textit{a macroscopic survival curve that deviates from a single exponential}.

\subsection{DDA-CFD Pipeline Framework for Airborne Virus Inactivation}

The resulting new DDA-CFD pipeline framework can now be summarized by:

\begin{equation}
	\mathbf{E}(\mathbf{r})
	\xrightarrow{\mathrm{DDA}}
	\xi(\mathbf{r})
	\xrightarrow{\text{Post-Process}}
	p_\xi(\xi)
	\xrightarrow{\text{Post-Process}}
    \mathcal{L}(s,d)
	\xrightarrow{\text{CFD}}
	\frac{N_s}{N_0}
	\label{eq:model_chain}
\end{equation}

And more explicitly by:

\begin{equation}
		\frac{N_s}{N_0}
		=
		\int_0^\infty
		e^{-Z H(t) \xi}
		p_\xi(\xi)
		\,d\xi
		=
		\mathcal{L}\{p_\xi\}(Z H(t))
	\label{eq:new_Ns}
\end{equation}

where $\xi$ is obtained from the DDA simulations; see eqn. (\ref{eq:xi}).

The above formulation framework therefore provides a multi-scale connection between the UV scattering problem and the virus inactivation response. 
The DDA simulation determines the spatial distribution of the UV field, while the susceptibility coefficient $Z$ describes the intrinsic response of the virus to UV radiation, and the local flow field and droplets dynamics will define the local UV exposure rate. A macroscopic survival fraction can then be obtained by averaging the local first-order virus inactivation kinetics over the spatially heterogeneous UV field inside each i-th droplet.

$N_s$ definition in eqn. (\ref{eq:new_Ns}) represents now an important more detailed version of the classical Chick--Watson law \cite{kowalski2009uvgi,Walker2007} for the dynamics of the number of survived virus copies $N_s(t)$. For the first time to our knowledge, eqn. (\ref{eq:new_Ns}) depends on both time and space. It takes into account the impact of UV light scattering on the total number of survived virus copies in airborne saliva droplets. 

\section{Laplace transform variation with droplet diameter, fluence and UV susceptibility}\label{parametric-fitting-of-the-laplace-transform}

\subsection{Definition of a fit model model for CFD applications}\label{motivation-and-definition-of-the-model}

Rewriting the Laplace transform again by adding its dependence on the saliva droplet diameter $d$ gives:
\begin{equation}
\mathcal{L}(s,d)
    = \int_{0}^{\infty} e^{-s\xi}p(\xi;d)\,\mathrm{d}\xi,
    \label{eq:laplace_definition}
\end{equation}    
    
\begin{equation}
\int_0^\infty p(\xi;d)\,\mathrm{d}\xi=1
\end{equation}

The dependence on \(d\) enters through the distribution of $\xi$. The objective is to represent the measured dependence of $\mathcal{L}$ on both $s$ and $d$ using a compact parametric expression to be applicable in CFD simulations.

Because $p(\xi;d)$ is a probability density supported on the non-negative real axis, its Laplace transform satisfies:

\begin{equation}
\mathcal{L}(0,d)=1,
    \qquad
    0\leq \mathcal{L}(s,d)\leq 1
\end{equation}
and is non-increasing with respect to $s$. 

When the relevant moments exist, its derivatives are related to the moments of the distribution:
\begin{equation}
(-1)^n\frac{\partial^n\mathcal{L}}{\partial s^n}(s,d)
    =\int_0^\infty \xi^n e^{-s\xi}p(\xi;d)\,\mathrm{d}\xi
    \geq 0
    \label{eq:complete_monotonicity}
\end{equation}  

In particular,
\begin{equation}
-\left.\frac{\partial\mathcal{L}}{\partial s}\right|_{s=0}
    =\mathbb{E}[\xi\mid d]
    \label{eq:mean_from_slope}
\end{equation}
of course, provided that the first moment is finite. 

The above approach provides useful guidelines to derive a fitting function of $\mathcal{L}(s,d)$ that will be implemented in the present CFD solver.

\subsection{Mathematical Derivation}

A convenient starting point is to assume that, for a fixed value of saliva droplet diameter $d$, the variable $\xi$ follows a distribution with shape
parameter $k>0$ and scale parameter $\theta(d)>0$ such that: 

\begin{equation}
p(\xi;d) = \frac{\xi^{k-1}e^{-\xi/\theta(d)}} {\Gamma(k)\theta(d)^k},
    \qquad \xi\geq 0
    \label{eq:gamma_pdf}
\end{equation}

where $\Gamma(k)$ denotes the distribution function. This density is normalized to unity and has mean $k\theta(d)$ and variance $k\theta(d)^2$.

Substituting eqn. (\ref{eq:gamma_pdf}) into the definition of the Laplace transform gives \begin{equation}
    \mathcal{L}(s,d) = \int_0^\infty e^{-s\xi}
       \frac{\xi^{k-1}e^{-\xi/\theta(d)}}
       {\Gamma(k)\theta(d)^k}\,\mathrm{d}\xi = \frac{1}{\Gamma(k)\theta(d)^k}
       \int_0^\infty \xi^{k-1}
       \exp\left[-\left(s+\frac{1}{\theta(d)}\right)\xi\right]
       \,\mathrm{d}\xi
\end{equation}

Using the Gamma-integral identity, this leads to:

\begin{equation}
\int_0^\infty x^{k-1}e^{-\lambda x}\,\mathrm{d}x
    = \frac{\Gamma(k)}{\lambda^k},
    \qquad k>0,\quad \lambda>0,
\end{equation}

and

\begin{equation}
\mathcal{L}(s,d)
    = \left[1+\theta(d)s\right]^{-k}
    \label{eq:gamma_laplace}
\end{equation}

To describe the observed dependence on \(d\) with a small number of parameters, the scale parameter is assumed to follow a power law such that:
\begin{equation}
\theta(d)=a\,d^{-\beta},
    \label{eq:theta_powerlaw}
\end{equation}
where \(a>0\) and \(\beta>0\). 

Combining eqn. (\ref{eq:gamma_laplace}) and eqn. \ref{eq:theta_powerlaw} yield a practical fitting model for UV scattering within CFD simulations such that:
\begin{equation}
\boxed{
    \mathcal{L}(s,d)
    = \left(1+\frac{as}{d^\beta}\right)^{-k}
    }
    \label{eq:fit_model}
\end{equation}

The fit model three parameters: $a$, $\beta$, and $k$ can thus be determined from DDA simulations. The shape parameter $k$ is assumed to be independent of $d$, while the scale of the distribution changes with $d$. 

In the present work, for $d \in [1-5]~\mu m$ and for all $s \leq 20$, we found that: $a=3.146$, $\beta=0.92$, $k=0.9324$ with overall accuracy of about $R^2 \approx 98 \%$.

\section{Scaling effect of UV scattering on virus survival in one single infected droplet}

\begin{figure}[ht]
    \includegraphics[width=\textwidth]{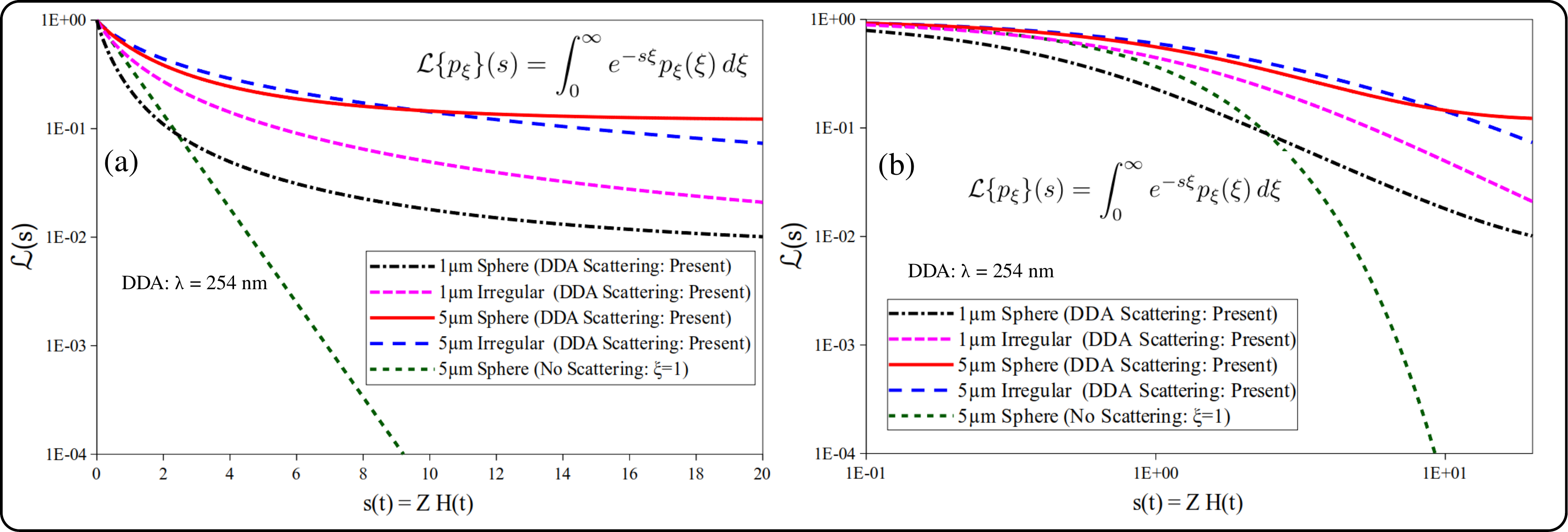}
    \caption{Scaling effect of UV scattering on virus survival fraction $N_s/N_0$ dynamics in one single infected dry droplet in still air UV-C irradiated at $\lambda=254~nm$. An example of the Laplace transform calculations: (a) $s$ in non-log-scale; (b) $s$ in log-scale. Simulations of UV scattering inside virus-infected droplet of two different effective sizes of 1 and 5 $\mu m$. Results for both spherical and irregular shaped dry saliva droplet and for both conditions of no scattering versus UV scattering where the latter is computed by the DDA method.}
    \label{fig:Laplace_transform_droplets}
\end{figure}

Figure \ref{fig:Laplace_transform_droplets} shows an example of the Laplace transform calculations, obtained from DDA simulations of UV scattering inside a single dry saliva droplet of both spherical and irregular shapes; see figure \ref{fig:DDA_p_of_xi_dried_vs_undried_saliva}. Results in figure \ref{fig:Laplace_transform_droplets} correspond to a saliva droplet of two different effective sizes $D_s=1 ~\mu m$ and $D_s=5 ~\mu m$, in still air (dry droplet with negligible evaporation, $\lambda=254~nm$ UV-C planar wave incidence from UV-C lamp with irradiance power $34.3~mW/cm^2$). Knowing that $\frac{N(t)}{N_0}=\mathcal{L}\{p_\xi\}(s)$, see eqn. (\ref{eq:new_Ns}), figure \ref{fig:Laplace_transform_droplets} clearly shows that assuming no scattering inside an infected droplet underestimates the prediction of survived virus copies for most values of $s(t)=ZH(t)$ especially at larger diameter values, nevertheless of the droplet's shape: e.g. spherical versus irregular shaped. Interestingly, it can be also observed that UV scattering computed by the DDA method, reveals a multi-exponent dynamic behavior of the total virus copies survival inside the droplet. 

\subsection{Validation of the DDA method applied to light scattering inside virions-infected saliva droplets}
\begin{figure}[ht]
    \includegraphics[width=\textwidth]{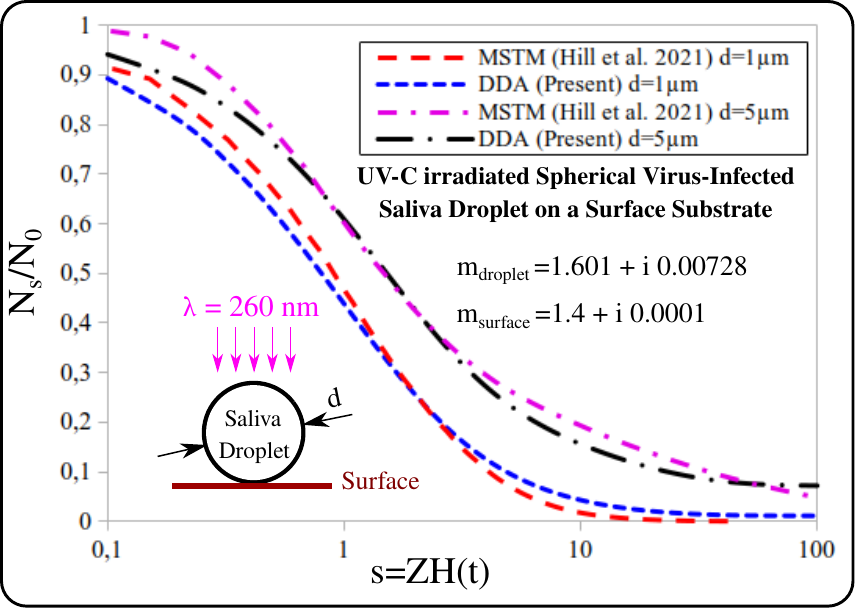}
    \caption{Scaling effect of UV scattering on virus survival fraction $N_s/N_0$ dynamics in one single infected dry droplet in still air UV-C irradiated at $\lambda=254~nm$. Results for two droplet diameters $d=1~\mu m$ and $d=5~\mu m$. An example of the Laplace transform calculations with $s$ in log scale and $N_s/N_0$ in normal scale. Simulations of UV scattering inside virus-infected droplet of two different effective sizes of 1 and 5 $\mu m$. Results for both spherical and irregular shaped dry saliva droplet and for both conditions of no scattering versus UV scattering where the latter is computed by the DDA method.}
    \label{fig:MSTM_DDA_compare_1um_5um_260nm}
\end{figure}

In order to investigate the validity of the DDA method applied to the UV-C irradiation ($\lambda=260~nm$) of virions inactivation in dry saliva droplets, figure \ref{fig:MSTM_DDA_compare_1um_5um_260nm} illustrates the DDA results applied to two dry $1 \mu m$ spherical droplets each placed on a surface substrate. These same two cases were also simulated by ~\citet{Hill2021,Doughty2021} but using the MSTM method for non-overlapping 12000 virions inside the saliva droplet. Figure \ref{fig:MSTM_DDA_compare_1um_5um_260nm} clearly shows that the DDA approach predicts a similar trend behavior of $N_s/N_0$ as function of $s=Z H(t)$ compared to that computed by ~\citet{Hill2021,Doughty2021}. The research works made by ~\citet{Hill2021,Doughty2021} employed the MSTM method for 12000 virions in saliva droplet placed on a surface substrate. The similarity observed in figure \ref{fig:MSTM_DDA_compare_1um_5um_260nm} between the Laplace behavior (thus $N_s/N_0$) of the DDA method and the absorption ration of the MSTM method ~\citet{Hill2021} reveals, for the first time, the validation of the DDA approach to virus inactivation in saliva droplets. Moreover, it is worth noting that the DDA method can treat different irregular shapes while the MSTM method is limited to spherical shapes only.

\section{CFD simulations results under neglected UV scattering}

\subsection{Validation of virus survival law coupling in CFD}

\begin{figure}[ht]
    \includegraphics[width=\textwidth]{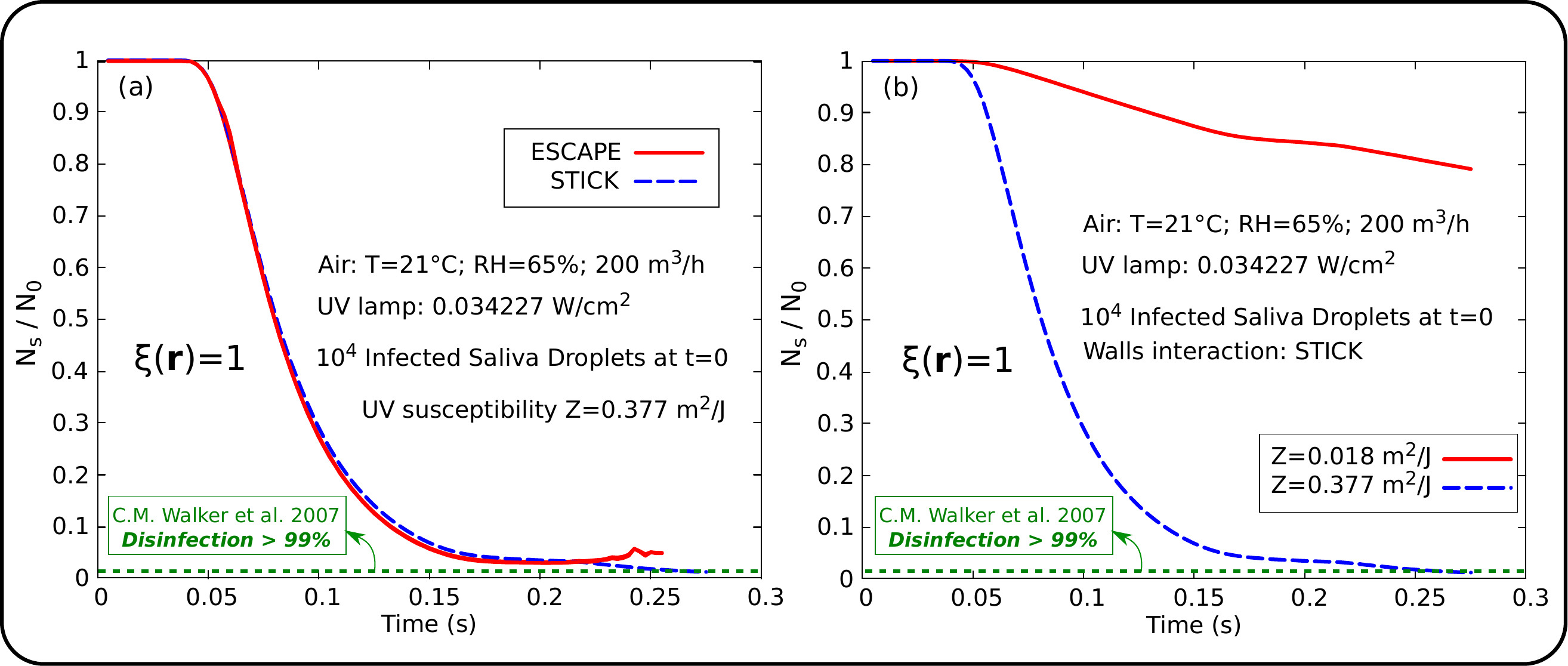}
    \caption{Effect of: (a) wall interaction law; and (b) UV Susceptibility constant Z; on the total fraction of survived viruses in airborne saliva droplets in present of evaporation. Case results for neglected UV scattering at $\xi(\mathbf{r})=1$ $\forall \mathbf{r}$. Case study scenario with initially emitted $10^4$ saliva droplets with effective size distribution $\qty{1}{\um} \leq d \leq \qty{500}{\um} $; see figure \ref{fig:PDF}-a. Worst case study scenario of air UV-purifier in terms of the highest studied inlet volumetric flow rate of \qty{200}{\m\cubed\per\hour}. Air temperature is \qty{21}{\degreeCelsius} initially with relative humidity RH=65\%. Cylindrical UV-C lamp steady surface temperature is \qty{38}{\degreeCelsius} with UV-C lamp irradiance power is \qty{34.3}{\mW\per\cm\squared}; see figure \ref{fig:BenchMarkDesign}. Note that the times at the curves ends, denotes the residence time before the airborne saliva droplets cloud leaves the computational domain of figure \ref{fig:BenchMarkDesign}.}
    \label{fig:results_200m3_plot_xi1}
\end{figure}

Figure \ref{fig:results_200m3_plot_xi1}-a shows the validation of the virus inactivation law implemented in the present CFD solver; e.g. compared to the experiments by Walker et al. \cite{Walker2007} who used a large value of $Z=\qty{0.377}{\m\squared\per\J}$. In figure \ref{fig:results_200m3_plot_xi1}-b, the virus survival behavior under a lower value of $Z=\qty{0.018}{\m\squared\per\J}$ is also shown. It is noteworthy that the lower value of $Z$ is usually obtained from continuum liquid experiments; see Kariwa et al. \cite{Kariwa2004}.

\subsection{Evaporation effect in CFD Modeling}

\begin{figure}[ht]
    \includegraphics[width=\textwidth]{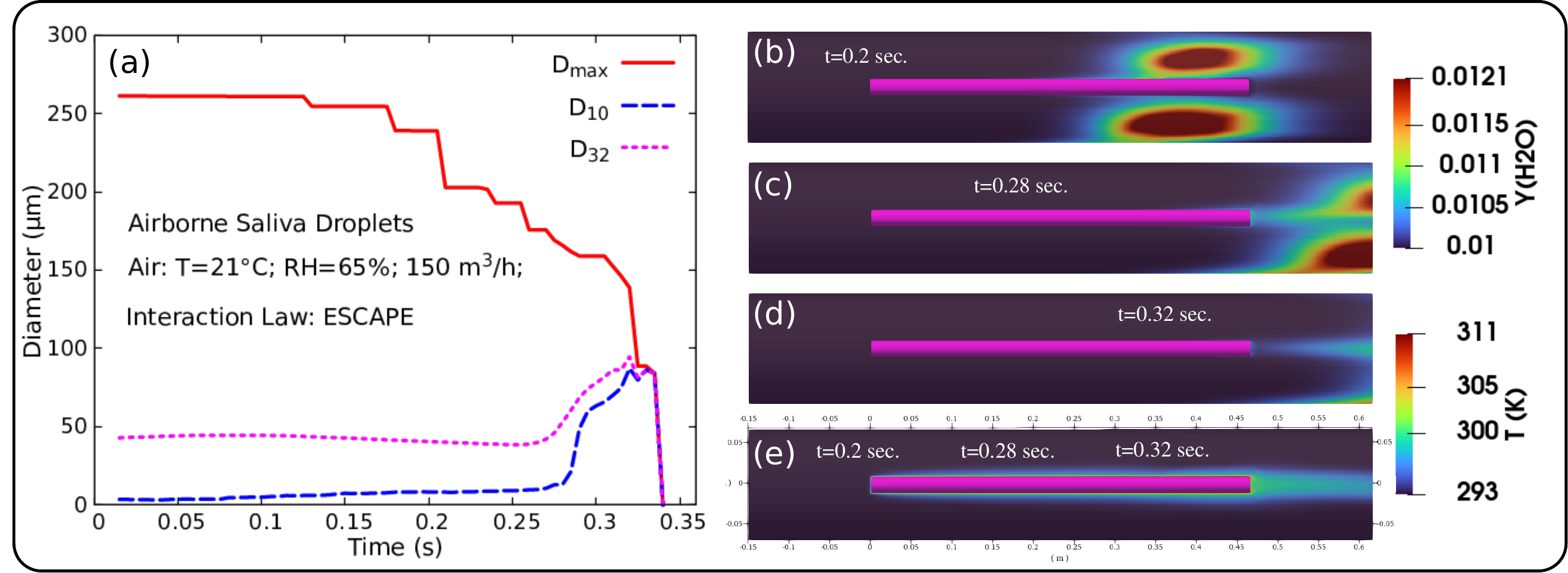}
    \caption{Effect of water content mass exchange between the saliva droplets and the surrounding air on statistical diameters. Case results for neglected UV scattering at $\xi(\mathbf{r})=1$ $\forall \mathbf{r}$. (a) statistical airborne saliva droplets diameters variation with time; Air mixture $Y_{H_2 O}$ at: (b) $t=\qty{0.2}{\s}$, (c) $t=\qty{0.28}{\s}$, (c) $t=\qty{0.32}{\s}$; (d) Air local Temperature field. Case study scenario with initially emitted $10^4$ saliva droplets with effective size distribution $\qty{1}{\um} \leq d \leq \qty{500}{\um} $; see figure \ref{fig:PDF}-a. Worst case study scenario of air UV-purifier in terms of the highest studied inlet volumetric flow rate of \qty{150}{\m\cubed\per\hour}. Air temperature is \qty{21}{\degreeCelsius} initially with relative humidity RH=65\%. Cylindrical UV lamp steady surface temperature is \qty{38}{\degreeCelsius} with UV irradiance power operating at \qty{34.3}{\mW\per\cm\squared}; see figure \ref{fig:BenchMarkDesign}. Results for an "ESCAPE" law for walls-particles interactions. The corresponding saliva droplets cloud dynamics were shown in figures \ref{fig:CFD1}, \ref{fig:CFD2} and \ref{fig:CFD3}.}
    \label{fig:plot_diameters_H2O_T_150m3}
\end{figure}

Figure \ref{fig:plot_diameters_H2O_T_150m3} shows the impact of heat and mass transfer, between the airborne saliva droplets and the surrounding air, on the statistical diameters $D_{10}$, $D_{32}$ and $D_{max}$ dynamics. Figure \ref{fig:plot_diameters_H2O_T_150m3} results correspond to a case study scenario with initially emitted $10^4$ saliva droplets with effective size distribution $\qty{1}{\um} \leq d \leq \qty{500}{\um}$ illustrated in figure \ref{fig:PDF}-a. Results are shown for \qty{150}{\m\cubed\per\hour} volumetric flow rate of air at temperature is \qty{21}{\degreeCelsius} and relative humidity of 65\% with \qty{38}{\degreeCelsius} of the UV lamp surface temperature (UV irradiance power of \qty{34.3}{\mW\per\cm\squared}, see figure \ref{fig:BenchMarkDesign}).
It can be clearly seen in figure \ref{fig:plot_diameters_H2O_T_150m3} that the maximum diameter $D_{max}$ is continuously decreasing which is explained by the evaporation of water content thus water mass loss from the droplets to the surrounding air. 
It can be also seen that $D_{32}$ slowly decreases between 0.05 and \qty{0.27}{\s} due to evaporation, but then increases at the end just before the last droplet leaves the domain with a residence time of about \qty{0.33}{\s}. This can be explained by the local variation of $Y_{H_2 O}$ and local temperature of air in the trailing edge zone of the UV lamp; e.g. zone where the droplets cloud becomes in the position after passing the UV lamp edge extremity in the flow direction (see Figures \ref{fig:plot_diameters_H2O_T_150m3}-c-d-e). The corresponding saliva droplets cloud dynamics were shown in figures \ref{fig:CFD1}, \ref{fig:CFD2} and \ref{fig:CFD3}.
Of course, one should keep in mind that all these coupled mechanisms that define the droplets size variation importantly depend on the local air speed, its local temperature, its local relative humidity, the surface temperature of the UV lamp's surface and the surrounding walls of the design; e.g. see figure \ref{fig:BenchMarkDesign}.

\subsection{Effect of initial saliva droplets size distribution, and air flow rate}

\begin{figure}[ht]
    \includegraphics[width=\textwidth]{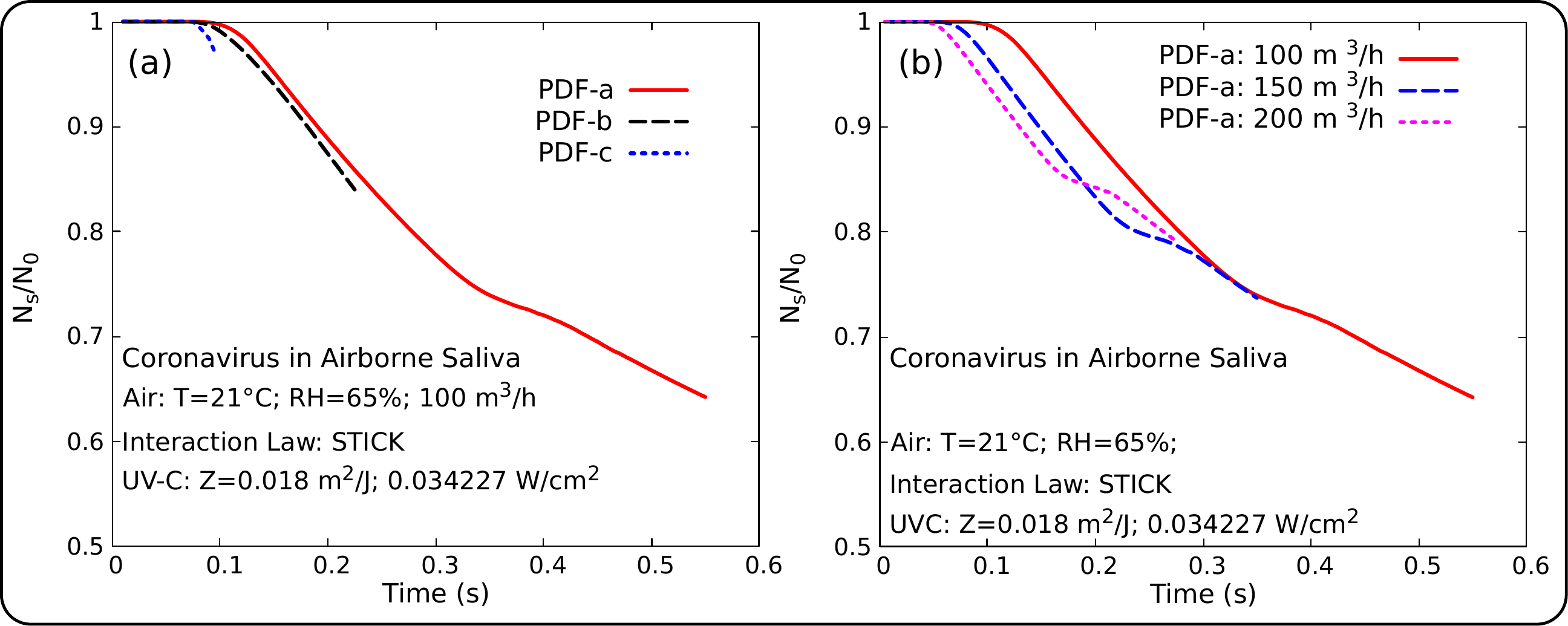}
    \caption{An example of the effects of: (a) saliva droplets initial size distribution, and (b) Air flow rates of \qty{100}{\m\cubed\per\hour}, \qty{150}{\m\cubed\per\hour} and \qty{200}{\m\cubed\per\hour} ; on the survival rate of airborne viruses. Case results for neglected UV scattering at $\xi(\mathbf{r})=1$ $\forall \mathbf{r}$. (a)-solid-line: initial PDF-a large size distribution, see figure \ref{fig:PDF}-a, with flow rate at \qty{100}{\m\cubed\per\hour}; (a)-dashed-line: initial PDF-b small size distribution, see figure \ref{fig:PDF}-b, with flow rate at \qty{100}{\m\cubed\per\hour}. Results for Initially $10^4$ airborne infected saliva droplets with neglected UV scattering at $\xi(\mathbf{r})=1$ $\forall \mathbf{r}$. UV-C planar wave at $\lambda=\qty{254}{\nm}$ with cylindrical UV-C lamp at \qty{34.3}{\mW\per\cm\squared}. Note that the times at the curves ends, denotes the residence time before the airborne saliva droplets cloud leaves the computational domain of figure \ref{fig:BenchMarkDesign}. Results for a "STICK" law for walls-particles interactions.} \label{fig:PDF_and_FlowRate_effects_10000p}
\end{figure}

For initially $10^4$ saliva droplets entering the air purifier design, it can be clearly seen in figure \ref{fig:PDF_and_FlowRate_effects_10000p}-a that in the case of a small PDF size distribution function (denoted PDF-b in figure \ref{fig:PDF}-b) the infected airborne saliva droplets cloud have lower residence time (about \qty{0.225}{\s}). At this PDF-b the virus-inactivated ratio $N_s/N_0 = 84\%$ is high even at relatively low air flow rate of \qty{100}{\m\cubed\per\hour}. Additionally, in the case of a larger PDF size distribution function (denoted PDF-a in figure \ref{fig:PDF}-a), the infected airborne saliva droplets cloud have higher residence time (about \qty{0.55}{\s}) where $N_s/N_0 = 65\%$ is also still high.
This highlights the fundamental importance of advanced CFD with multiphysics multiscale modeling in predicting the inactivation of airborne viruses, which are usually very difficult and very dangerous to quantify experimentally.
It can be also observed in figure \ref{fig:PDF_and_FlowRate_effects_10000p}-b that increasing the flow rate between \qty{100}{\m\cubed\per\hour} and \qty{200}{\m\cubed\per\hour}, does not induce a linear variation in the total residence time of the airborne saliva droplets. This can be explained by the complex coupling between local air flow dynamics and heat and mass transfer local mechanisms. This also highlights that an optimal operating flow rate value will be interesting to find; e.g. depending on the design of the air purifier disinfection system. This also highlights the fundamental role of employing advanced CFD with enriched multiphysics multiscale modeling in accurately predicting the inactivation of airborne viruses, which are usually difficult/dangerous to be quantified/manipulated experimentally.

\subsection{Effect of initially emitted saliva droplets number}

\begin{figure}[ht]
    \centering
    \includegraphics[width=0.7\textwidth]{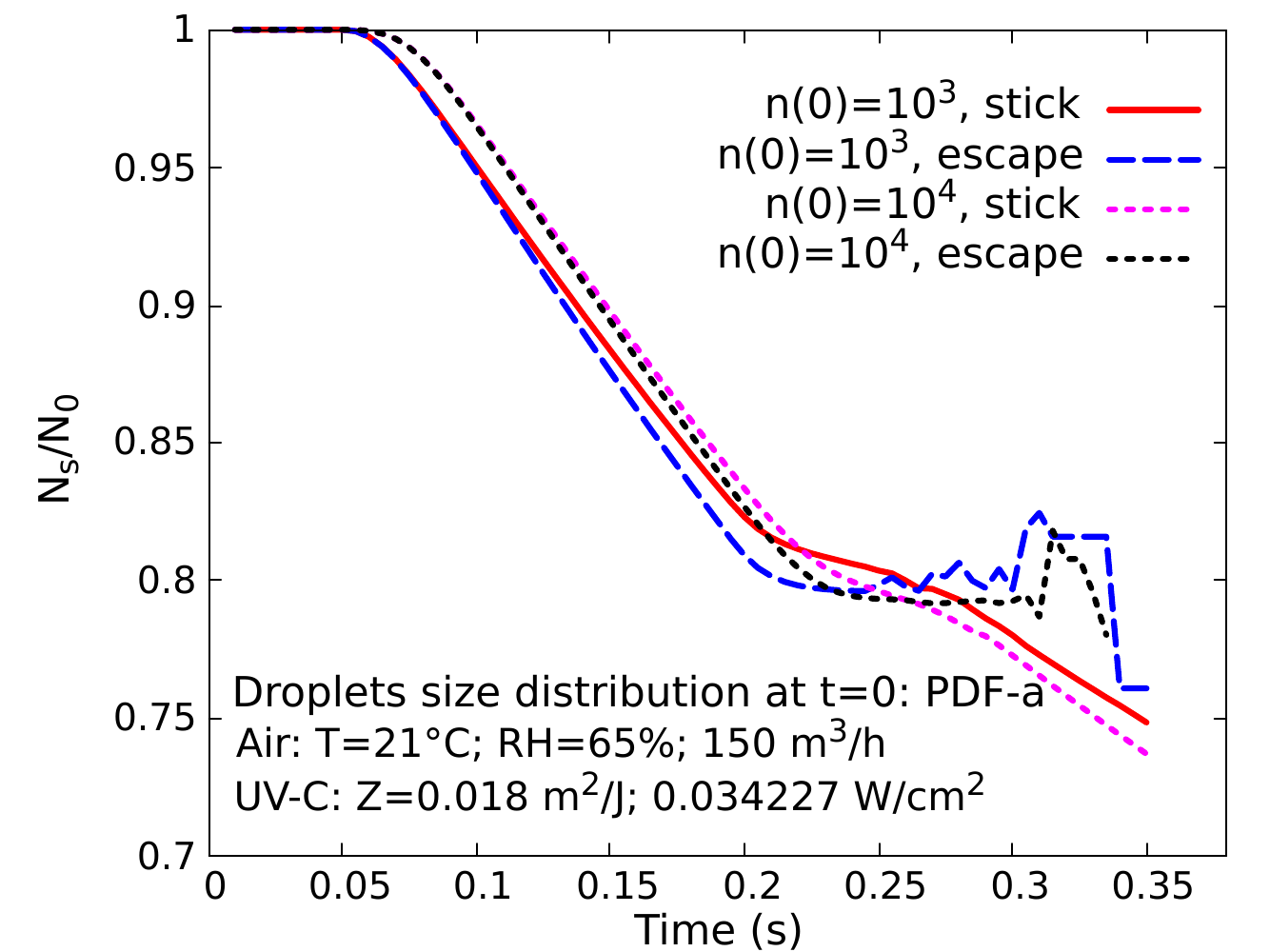}
    \caption{Effect of Initial Saliva Droplets Number on the Survival of Airborne Viruses; including evaporation. Case results for neglected UV scattering at $\xi(\mathbf{r})=1$ $\forall \mathbf{r}$. Results at $Z=\qty{0.018}{\m\squared\per\J}$ and for two walls-particles interactions laws "stick" versus "escape" and for two different initial number of droplets $10^3$ and $10^4$ at $t=0$. Results for initially PFD-a large droplets size distribution, see figure \ref{fig:PDF}-a, in an air flow at \qty{150}{\m\cubed\per\hour} at \qty{21}{\degreeCelsius} and RH=65\% with UV lamp surface temperature of \qty{38}{\degreeCelsius}; see figure \ref{fig:BenchMarkDesign}. Results for UV-C planar wave at $\lambda=\qty{254}{\nm}$ with cylindrical UV-C lamp at \qty{34.3}{\mW\per\cm\squared}.}
    \label{fig:stick_escape_150m3_plot_Particles_Number_Effect}
\end{figure}

Figure \ref{fig:stick_escape_150m3_plot_Particles_Number_Effect} clearly illustrates the effect of the initial saliva droplets number $n(0)$ on the survival of airborne viruses in saliva in the case of an intermediate air flow rate of \qty{150}{\m\cubed\per\hour}. It can be clearly observed that the initially infected saliva droplets total number at inlet ($10^3$ versus $10^4$), and the walls-droplets interaction law (\textit{stick} versus \textit{escape}) both affect the virus survival rate, even at low $Z$ value. In other words, this also affects the overall efficiency of the air purifier disinfection system represented by $N_s/N_0$ in figure \ref{fig:stick_escape_150m3_plot_Particles_Number_Effect}.

\newpage
\pagebreak

\section{CFD simulations results under non-neglected UV scattering}

Taking into account in the CFD solver the effect of non-neglected UV scattering on virus survival, the CFD-DDA approach is applied for the first time to our knowledge, under the following scenario conditions:
\begin{itemize}
    \item Small number of initially 1000 airborne saliva droplets ($n(0)=10^3$) randomly entering the inlet of the purifier system design shown in figure \ref{fig:BenchMarkDesign},
    \item Negligible shadowing is assumed between the airborne droplets, 
    \item Maximum indoor air circulation flow rate $200~m^3/h$ versus a minimum indoor air circulation flow rate $100~m^3/h$, 
    \item High virus UV susceptibility constant $Z=\qty{0.377}{\m\squared\per\J}$, and
    \item Dried saliva droplets of initial Rosin--Rammler small size distribution between 1 and 5 $\mu~m$ (eqn.\ref{eq:PDF} with parameters: $\beta=3.5~µm$, $n=2$, $d_{min}=1~\mu m$, $d_{min}=5~\mu m$), see figure \ref{fig:PDF}; e.g. assuming infected saliva droplets dried before entering the inlet of purifier; see figure \ref{fig:Evaporation_COV_UV}.
\end{itemize}

The above scenario conditions allows one to clearly investigate the impact of UV scattering intensity ($\xi(\mathbf{r}) \neq1$) in CFD as represented by the distribution functions $p(\xi)$; e.g. see figure \ref{fig:irregularDropletGeometry} and eqn. \ref{eq:new_Ns}.

\begin{figure}[ht]
    \centering
    \includegraphics[width=\textwidth]{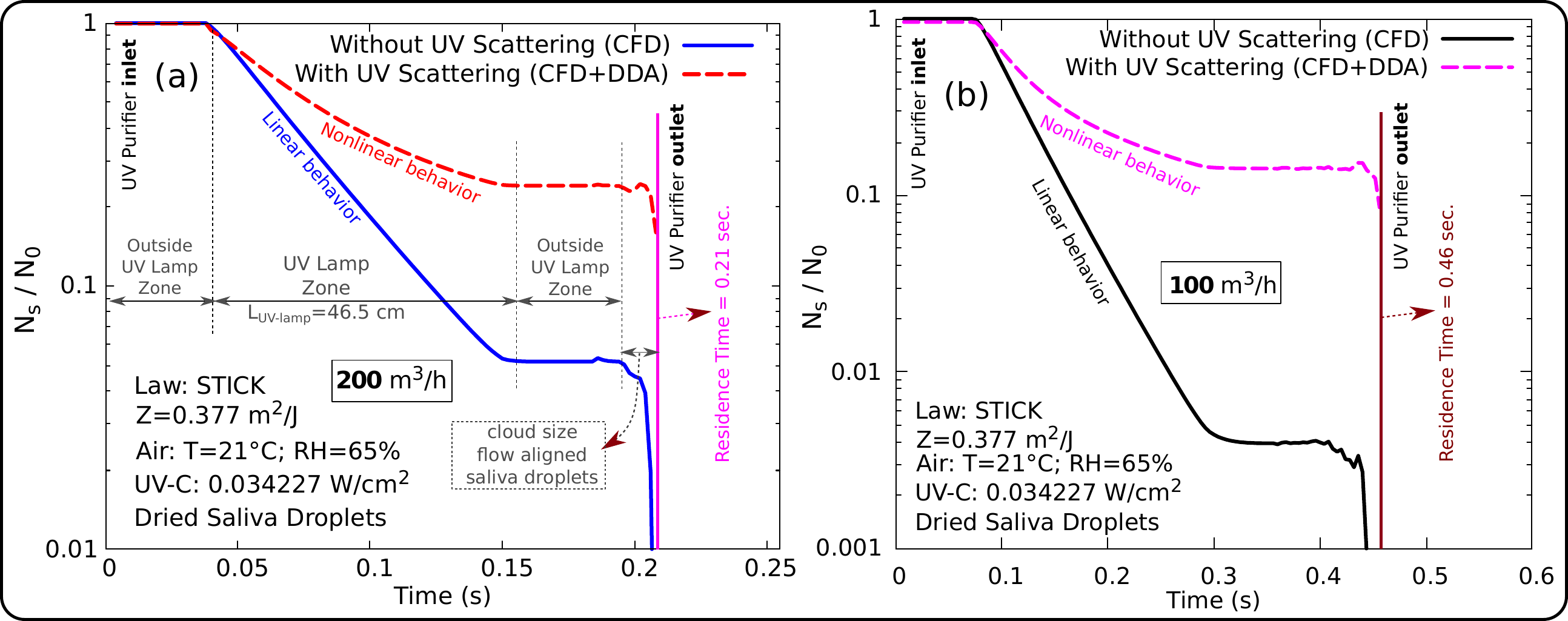}
    \caption{Effect of UV scattering on the survival of viruses in infected airborne spherical dry saliva droplets of initial size distribution: $1~\mu m\leq d~\leq 5~\mu m$ (PDF-c). (a) airflow rate at $200~m^3/h$; (b) airflow rate at $100~m^3/h$. (dashed-line) Employing both advanced CFD+DDA simulations the UV scattering phenomena induce a non-linear multi-exponent virus survival dynamics behavior. (solid-line) Neglecting the UV scattering in CFD can highly underestimate the total virus copies survival in infected dried airborne saliva droplets. Results at $200~m^3/h$, $T_{air}=21^{\circ}$C, RH=65\% and UV lamp power of 10 Watts corresponding to $34.3~mW/cm^2$; see the benchmark design of the UV-based air-purifier in figure \ref{fig:BenchMarkDesign}.}
    \label{fig:UV_scattering_effect_200_100_m3_1000p}
\end{figure}

Figure \ref{fig:UV_scattering_effect_200_100_m3_1000p} shows clearly the effect of UV scattering on the survival of viruses in infected dried airborne saliva droplets ($1~\mu m\leq d~\leq 5~\mu m$) under airflow rates of $200~m^3/h$ (figure \ref{fig:UV_scattering_effect_200_100_m3_1000p}-a) and $100~m^3/h$ (figure \ref{fig:UV_scattering_effect_200_100_m3_1000p}-b). When DDA UV scattering phenomenon is considered inside the CFD simulations (see red dashed-line in figure \ref{fig:UV_scattering_effect_200_100_m3_1000p}-a), the virus survival dynamic witness a multi-exponent non-linear behavior. However, under neglected UV scattering in CFD (see blue solid-line in figure \ref{fig:UV_scattering_effect_200_100_m3_1000p}-b), a classical one-exponent virus survival dynamic behavior is observed.
These results highlight very clearly the important role of coupling between DDA and CFD simulations for more accurate predictions of airborne virus survival dynamics, especially when designing flow circulation UV-based airborne virus inactivation systems or air purifiers disinfection equipment.

\section{Conclusion and perspectives}

The following major fundamental conclusions can be drawn:

\begin{enumerate}
    \item Neglecting UV scattering in CFD modeling can dangerously underestimate the total number $N_s$ of survived virion copies in airborne saliva droplets, even at very high values of airborne viruses UV susceptibility of the order of $Z=\qty{0.377}{\m\squared\per\J}$, and even at high values of airflow rate of the order of $200~m^3/h$. 
    \item UV light scattering phenomenon thus can not be neglected in advanced CFD modeling and simulation of airborne virus purifier and disinfection systems.
    \item DDA results used to develop an enhanced law for virus inactivation. It provides a more detailed version of the classical Chick--Watson law in order to account for UV scattering phenomena in both spherical and non-spherical airborne infected saliva droplets.
    \item The evaporation process of airborne saliva droplets impacts their regular shape which also induces a vital impact on the UV scattering field. Drying process induces larger refractive index and thus potentially greater shielding of embedded organisms in the airborne dehydrated saliva droplets.
    \item Even very low values of virus viral loads (\qty{100}{\nm} in diameter) in infected airborne saliva droplets, can be as dangerous as larger values within the whole range of the size distribution of expelled airborne human saliva droplets. This is due to UV scattering phenomenon accompanied with negligible diffusion of virion inclusions especially inside dehydrated saliva droplets compared to the UV equipment residence time scales.    
    \item The initial emitted saliva droplet size distribution (e.g. small, versus large effective diameters and their polydispersity degree) will importantly affect the survival of airborne viruses in dynamic systems. 
    \item Without developing advanced multiphysics multiscale enriched models in CFD, UV-C air purifiers disinfection systems designed for airborne droplets large residence times will not guarantee that the virus inclusions in all the airborne saliva droplets will be inactivated by the UV-C irradiance field.
    \item Due to the public health risks associated with airborne viruses manipulation in strict research laboratory policies and risk associated regulations, experimental setups and measurements are still very limited to quantify a precised susceptibility and precised UV dose for airborne virus disinfection. Thus advanced CFD becomes an extremely useful tool in this direction.
\end{enumerate}

The above conclusions open new perspectives and guidelines as the following:

\begin{itemize}
    \item Conduct new DDA simulations for different angle orientations of a irregular shaped dried saliva droplets with different sizes and at different refraction indices properties,
    \item Conduct new CFD simulations and seek new designs/mechanisms that not only increase the residence time but also that allow for a local spin/rotation of the infected saliva microdroplets,
    \item Seek new approaches to enhance the local mixing of the airborne droplets under flow conditions,
    \item Develop advanced computations at the scale of a single infected saliva droplet to better understand the fluid-microdroplet-interactions in presence of virion nano-inclusions under different conditions.
\end{itemize}

\section*{Author contributions}

\textbf{T.D.}: Conceptualization, Methodology, Analysis, Software (CFD), CFD-DDA Coupling, Visualization, Original draft. \textbf{M.A.Y.}: Methodology, Analysis, Software (DDA), Review and Editing.

\section*{Conflict of Interest}

The authors have no conflicts to disclose.

\section*{Funding}

M.Y. acknowledges the support from the Normandy Region through project RADDAERO. T.D. acknowledges the support from CRIANN datacenter at the Normandy Region under project no. 2025011.

\section*{Data availability} 

The data that support the findings of this study are available upon reasonable request from the corresponding authors.

\bibliographystyle{unsrtnat}
\bibliography{aipRef.bib}

@article{Longest2024,
    author = {Longest, Alexandra K. and Rockey, Nicole C. and Lakdawala, Seema S. and Marr, Linsey C.},
    title = {Review of factors affecting virus inactivation in aerosols and droplets},
    journal = {Journal of The Royal Society Interface},
    volume = {21},
    number = {215},
    pages = {20240018},
    year = {2024},
    month = {06},
    issn = {1742-5689},
    doi = {10.1098/rsif.2024.0018},
    url = {https://doi.org/10.1098/rsif.2024.0018},
}

@article{Sapkota2026,
author = {Deepak Sapkota and Yuhui Guo and Amrita Chakraborty and James Hu and Harris Xie and Ian Wu and Moon J. Kim and Hui Ouyang},
title = {Organic components modulate the morphology of respirable aerovirology-relevant aerosols},
journal = {Aerosol Science and Technology},
volume = {0},
number = {0},
pages = {1--14},
year = {2026},
publisher = {Taylor \& Francis},
doi = {10.1080/02786826.2026.2637801},
URL = {https://doi.org/10.1080/02786826.2026.2637801},
}

@article{McNeely2021,
  author  = {McNeely, Jeffrey A.},
  title   = {Nature and COVID-19: The pandemic, the environment, and the way ahead},
  journal = {Ambio},
  year    = {2021},
  volume  = {50},
  number  = {4},
  pages   = {767--781},
  doi     = {10.1007/s13280-020-01447-0},
  url     = {https://doi.org/10.1007/s13280-020-01447-0}
}

@article{Seitz2020,
  author    = {Claudia, Seitz},
  title     = {Genetic Material and Sequence Data to Protect Global Health in the Light of Pandemic Outbreaks: Mapping the Legal Landscape under European and International Law},
  journal   = {European Journal of Health Law},
  volume    = {27},
  number    = {3},
  pages     = {232--241},
  year      = {2020},
  doi       = {10.1163/15718093-BJA10014}
}

@article{Dbouk2021b,
    author = {Dbouk, Talib and Drikakis, Dimitris},
    title = {Fluid dynamics and epidemiology: Seasonality and transmission
          dynamics},
    journal = {Physics of Fluids},
    volume = {33},
    number = {2},
    pages = {021901},
    year = {2021},
    month = {02},
    issn = {1070-6631},
    doi = {10.1063/5.0037640},
    url = {https://doi.org/10.1063/5.0037640},
}

@article{Dbouk2021c,
    author = {Dbouk, Talib and Drikakis, Dimitris},
    title = {On pollen and airborne virus transmission},
    journal = {Physics of Fluids},
    volume = {33},
    number = {6},
    pages = {063313},
    year = {2021},
    month = {06},
    issn = {1070-6631},
    doi = {10.1063/5.0055845},
    url = {https://doi.org/10.1063/5.0055845},
}

@article{Wyllie2020,
  author = {Anne L. Wyllie  and John Fournier  and Arnau Casanovas-Massana  and Melissa Campbell  and Maria Tokuyama  and Pavithra Vijayakumar  and Joshua L. Warren  and Bertie Geng  and M. Catherine Muenker  and Adam J. Moore  and Chantal B.F. Vogels  and Mary E. Petrone  and Isabel M. Ott  and Peiwen Lu  and Arvind Venkataraman  and Alice Lu-Culligan  and Jonathan Klein  and Rebecca Earnest  and Michael Simonov  and Rupak Datta  and Ryan Handoko  and Nida Naushad  and Lorenzo R. Sewanan  and Jordan Valdez  and Elizabeth B. White  and Sarah Lapidus  and Chaney C. Kalinich  and Xiaodong Jiang  and Daniel J. Kim  and Eriko Kudo  and Melissa Linehan  and Tianyang Mao  and Miyu Moriyama  and Ji E. Oh  and Annsea Park  and Julio Silva  and Eric Song  and Takehiro Takahashi  and Manabu Taura  and Orr-El Weizman  and Patrick Wong  and Yexin Yang  and Santos Bermejo  and Camila D. Odio  and Saad B. Omer  and Charles S. Dela Cruz  and Shelli Farhadian  and Richard A. Martinello  and Akiko Iwasaki  and Nathan D. Grubaugh  and Albert I. Ko },
  title = {Saliva or Nasopharyngeal Swab Specimens for Detection of SARS-CoV-2},
  journal = {New England Journal of Medicine},
  volume = {383},
  number = {13},
  pages = {1283-1286},
  year = {2020},
  doi = {10.1056/NEJMc2016359},
  URL = {https://www.nejm.org/doi/full/10.1056/NEJMc2016359}
}

@article{carpenter2013salivary,
  author  = {Carpenter, G. H.},
  title   = {The secretion, components, and properties of saliva},
  journal = {Annual Review of Food Science and Technology},
  volume  = {4},
  pages   = {267--276},
  year    = {2013},
  doi     = {10.1146/annurev-food-030212-182700}
}

@article{humphrey2001role,
  author  = {Humphrey, S. P. and Williamson, R. T.},
  title   = {A review of saliva: Normal composition, flow, and function},
  journal = {Journal of Prosthetic Dentistry},
  volume  = {85},
  number  = {2},
  pages   = {162--169},
  year    = {2001},
  doi     = {10.1067/mpr.2001.113778}
}

@Article{Trivellin2021,
author = {Trivellin, Nicola and Piva, Francesco and Fiorimonte, Davide and Buffolo, Matteo and De Santi, Carlo and Orlandi, Viviana Teresa and Dughiero, Fabrizio and Meneghesso, Gaudenzio and Zanoni, Enrico and Meneghini, Matteo},
title = {UV-Based Technologies for SARS-CoV2 Inactivation: Status and Perspectives},
journal = {Electronics},
volume = {10},
year = {2021},
number = {14},
url = {https://www.mdpi.com/2079-9292/10/14/1703},
issn = {2079-9292},
DOI = {10.3390/electronics10141703}
}

@article{Marshall2024,  
author = {S. Marshall and R. Basile and D. Tanke and N. Francois and P. Nekolny and F. Duchaine and S. Sankurantripati},
title = {Ultraviolet Germicidal Irradiation Development Method for Transportation Disinfection Modelling},
url={https://emas.nafems.org/index.php/v1/article/view/7},  
doi={10.59972/zsj0dnh6}, 
journal={Engineering Modelling, Analysis and Simulation}, 
year={2024}, 
month={8},
volume={1}, 
}

@article{SANKURANTRIPATI2025,
title = {Large eddy simulations to investigate airborne virus inactivation using a ultraviolet air purifier with Lagrangian tracking},
journal = {Journal of Aerosol Science},
volume = {189},
pages = {106642},
year = {2025},
issn = {0021-8502},
doi = {https://doi.org/10.1016/j.jaerosci.2025.106642},
url = {https://www.sciencedirect.com/science/article/pii/S0021850225001193},
author = {S. Sankurantripati and F. Duchaine and N. Francois and S. Marshall and P. Nekolny}
}

@article{Beggs2020,
  author  = {Beggs, Clive B. and Avital, Eli J.},
  title   = {Upper-room ultraviolet air disinfection might help to reduce COVID-19 transmission in buildings: A feasibility study},
  journal = {PeerJ},
  year    = {2020},
  volume  = {8},
  pages   = {e10196},
  doi     = {10.7717/peerj.10196},
  url     = {https://doi.org/10.7717/peerj.10196}
}

@article{Kariwa2004,
    author = {H. Kariwa and N. Fujii and I. Takashima},
    title = {Inactivation of SARS coronavirus by means of povidone-iodine, physical conditions, and chemical reagents},
    journal = {Jpn J Vet Res.},
    year = {2004},
    volume = {52},
    number = {3},
    pages = {105--12}
}

@article{DARNELL2004,
title = {Inactivation of the coronavirus that induces severe acute respiratory syndrome, SARS-CoV},
journal = {Journal of Virological Methods},
volume = {121},
number = {1},
pages = {85-91},
year = {2004},
issn = {0166-0934},
doi = {https://doi.org/10.1016/j.jviromet.2004.06.006},
url = {https://www.sciencedirect.com/science/article/pii/S016609340400179X},
author = {Miriam E.R. Darnell and Kanta Subbarao and Stephen M. Feinstone and Deborah R. Taylor}
}

@article{Jensen1964,
author = {Marcus M. Jensen},
title = {Inactivation of Airborne Viruses by Ultraviolet Irradiation},
journal = {Applied Microbiology},
volume = {12},
number = {5},
pages = {418-420},
year = {1964},
doi = {10.1128/am.12.5.418-420.1964},
URL = {https://journals.asm.org/doi/abs/10.1128/am.12.5.418-420.1964}
}

@article{Walker2007,
author = {Walker, Christopher M. and Ko, GwangPyo},
title = {Effect of Ultraviolet Germicidal Irradiation on Viral Aerosols},
journal = {Environmental Science \& Technology},
volume = {41},
number = {15},
pages = {5460-5465},
year = {2007},
doi = {10.1021/es070056u},
note ={PMID: 17822117},
URL = {https://doi.org/10.1021/es070056u},
}

@article{Bedell2016,
  author  = {Bedell, Kurt and Buchaklian, Albert H. and Perlman, Stanley},
  title   = {Efficacy of an Automated Multiple Emitter Whole-Room Ultraviolet-C Disinfection System Against Coronaviruses MHV and MERS-CoV},
  journal = {Infection Control \& Hospital Epidemiology},
  year    = {2016},
  volume  = {37},
  number  = {5},
  pages   = {598--599},
  doi     = {10.1017/ice.2015.348},
  url     = {https://doi.org/10.1017/ice.2015.348}
}

@article{Welch2018,
  author  = {Welch, David and Buonanno, Manuela and Grilj, Vanja and 
             Shuryak, Igor and Crickmore, Curtis and Bigelow, Ann W. and 
             Randers-Pehrson, Gerhard and Johnson, Gregory W. and Brenner, David J.},
  title   = {Far-UVC light: A new tool to control the spread of airborne-mediated microbial diseases},
  journal = {Scientific Reports},
  year    = {2018},
  volume  = {8},
  number  = {1},
  pages   = {2752},
  doi     = {10.1038/s41598-018-21058-w},
  url     = {https://doi.org/10.1038/s41598-018-21058-w}
}

@article{Han2013,
    author = {Han, Z. Y. and Weng, W. G. and Huang, Q. Y.},
    title = {Characterizations of particle size distribution of the droplets exhaled by sneeze},
    journal = {Journal of The Royal Society Interface},
    volume = {10},
    number = {88},
    pages = {20130560},
    year = {2013},
    month = {11},
    issn = {1742-5689},
    doi = {10.1098/rsif.2013.0560},
    url = {https://doi.org/10.1098/rsif.2013.0560},
}

@article{Xie:2009,
    author = {Xie, Xiaojian and Li, Yuguo and Sun, Hequan and Liu, Li},
    title = {Exhaled droplets due to talking and coughing},
    journal = {Journal of The Royal Society Interface},
    volume = {6},
    number = {6},
    pages = {S703-S714},
    year = {2009},
    month = {10},
    issn = {1742-5689},
    doi = {10.1098/rsif.2009.0388.focus},
}

@article{Asadi2020VoicingAerosol,
  author  = {Asadi, Sima and Wexler, Anthony S. and Cappa, Christopher D. and Barreda, Santiago and Bouvier, Nicole M. and Ristenpart, William D.},
  title   = {Effect of voicing and articulation manner on aerosol particle emission during human speech},
  journal = {PLOS ONE},
  year    = {2020},
  volume  = {15},
  number  = {1},
  pages   = {e0227699},
  doi     = {10.1371/journal.pone.0227699},
  pmid    = {31986165},
  url     = {https://doi.org/10.1371/journal.pone.0227699}
}

@article{Stadnytskyi2020AirborneLifetime,
  author  = {Stadnytskyi, Valentyn and Bax, Christina E. and Bax, Adriaan and Anfinrud, Philip},
  title   = {The airborne lifetime of small speech droplets and their potential importance in SARS-CoV-2 transmission},
  journal = {Proceedings of the National Academy of Sciences},
  year    = {2020},
  volume  = {117},
  number  = {22},
  pages   = {11875--11877},
  doi     = {10.1073/pnas.2006874117},
  pmid    = {32404416},
  url     = {https://doi.org/10.1073/pnas.2006874117}
}

@article{Alic2021ThreeStageInactivation,
  author  = {Alic, Faruk},
  title   = {Analytical modeling of three-stage inactivation of viruses within droplets and solid porous particles},
  journal = {European Physical Journal Plus},
  year    = {2021},
  volume  = {136},
  number  = {6},
  pages   = {663},
  doi     = {10.1140/epjp/s13360-021-01651-1},
  url     = {https://doi.org/10.1140/epjp/s13360-021-01651-1}
}

@article{BenMa2021,
author = {Ben Ma and Patricia M. Gundy and Charles P. Gerba and Mark D. Sobsey and Karl G. Linden},
title = {UV Inactivation of SARS-CoV-2 across the UVC Spectrum: KrCl* Excimer, Mercury-Vapor, and Light-Emitting-Diode (LED) Sources},
journal = {Applied and Environmental Microbiology},
volume = {87},
number = {22},
pages = {e01532-21},
year = {2021},
doi = {10.1128/AEM.01532-21},
URL = {https://journals.asm.org/doi/abs/10.1128/aem.01532-21}
}

@article{Rayleigh1871SkyLight,
  author  = {Rayleigh, Lord},
  title   = {On the light from the sky, its polarization and colour},
  journal = {Philosophical Magazine},
  year    = {1871},
  volume  = {41},
  pages   = {107--120},
  doi     = {10.1080/14786447108640452}
}

@book{Bohren1983AbsorptionScattering,
  author    = {Bohren, Craig F. and Huffman, Donald R.},
  title     = {Absorption and Scattering of Light by Small Particles},
  publisher = {Wiley},
  year      = {1983}
}

@article{Mie1908TrubeMedien,
  author  = {Mie, Gustav},
  title   = {Beiträge zur Optik trüber Medien, speziell kolloidaler Metallösungen},
  journal = {Annalen der Physik},
  year    = {1908},
  volume  = {330},
  number  = {3},
  pages   = {377--445},
  doi     = {10.1002/andp.19083300302}
}

@article{Dombrovsky2003SpectralDieselDroplets,
  author  = {Dombrovsky, Leonid A. and Sazhin, Sergei S. and Mikhalovsky, Sergey V. and Wood, R. and Heikal, Morgan R.},
  title   = {Spectral properties of diesel fuel droplets},
  journal = {Fuel},
  year    = {2003},
  volume  = {82},
  number  = {1},
  pages   = {15--22},
  doi     = {10.1016/S0016-2361(02)00200-4},
  url     = {https://doi.org/10.1016/S0016-2361(02)00200-4}
}

@article{Dombrovsky2003SemitransparentDroplet,
  author  = {Dombrovsky, Leonid A. and Sazhin, Sergei S.},
  title   = {Absorption of thermal radiation in a semi-transparent spherical droplet: A simplified model},
  journal = {International Journal of Heat and Fluid Flow},
  year    = {2003},
  volume  = {24},
  number  = {6},
  pages   = {919--927},
  doi     = {10.1016/S0142-727X(03)00084-5},
  url     = {https://doi.org/10.1016/S0142-727X(03)00084-5}
}

@article{Yurkin2007DDAReview,
  author  = {Yurkin, Maxim A. and Hoekstra, Alfons G.},
  title   = {The Discrete Dipole Approximation: An Overview and Recent Developments},
  journal = {Journal of Quantitative Spectroscopy and Radiative Transfer},
  year    = {2007},
  volume  = {106},
  number  = {1--3},
  pages   = {558--589},
  doi     = {10.1016/j.jqsrt.2007.01.034},
  url     = {https://doi.org/10.1016/j.jqsrt.2007.01.034}
}

@incollection{Yurkin2023Chapter,
	address = {Amsterdam},
	title = {Discrete dipole approximation},
	copyright = {All rights reserved},
	isbn = {978-0-323-99901-4},
	url = {https://doi.org/10.1016/B978-0-323-99901-4.00020-2},
	doi = {10.1016/B978-0-323-99901-4.00020-2},
	booktitle = {Light, {Plasmonics} and {Particles}},
	publisher = {Elsevier},
	author = {Yurkin, M. A.},
	editor = {Menguc, M. P. and Francoeur, M.},
	year = {2023},
	pages = {167--198}
}

@article{Yurkin2011ADDA,
	title = {The discrete-dipole-approximation code {ADDA}: {Capabilities} and known limitations},
	volume = {112},
	copyright = {All rights reserved},
	issn = {0022-4073},
	shorttitle = {The discrete-dipole-approximation code {ADDA}},
	doi = {10.1016/j.jqsrt.2011.01.031},
	number = {13},
	urldate = {2011-02-07},
	journal = {J. Quant. Spectrosc. Radiat. Transfer},
	author = {Yurkin, Maxim A. and Hoekstra, Alfons G.},
	month = sep,
	year = {2011},
	pages = {2234--2247}
}

@article{Yurkin2010FCD,
	title = {Application of the discrete dipole approximation to very large refractive indices: {Filtered} coupled dipoles revived},
	volume = {82},
	copyright = {All rights reserved},
	url = {http://pre.aps.org/abstract/PRE/v82/i3/e036703},
	doi = {10.1103/PhysRevE.82.036703},
	number = {3},
	journal = {Phys. Rev. E},
	author = {Yurkin, M. A. and Min, M. and Hoekstra, A. G.},
	year = {2010},
	pages = {036703}
}

@inproceedings{schuh2007pip,
	address = {Bremen, Germany},
	title = {Arbitrary particle shape modeling in {DDSCAT} and validation of simulation results},
	url = {https://pdfs.semanticscholar.org/0db5/4bdd468a5daaa389e21485eacc5a481f8c8e.pdf},
	booktitle = {Proceedings of the {DDA}-{Workshop}},
	author = {Schuh, R.},
	editor = {Wriedt, T. and Hoekstra, A. G.},
	month = mar,
	year = {2007},
	pages = {22--24},
}

@article{Dbouk2020a,
    author = {Dbouk, Talib and Drikakis, Dimitris},
    title = {On coughing and airborne droplet transmission to humans},
    journal = {Physics of Fluids},
    volume = {32},
    number = {5},
    pages = {053310},
    year = {2020},
    month = {05},
    issn = {1070-6631},
    doi = {10.1063/5.0011960},
    url = {https://doi.org/10.1063/5.0011960},
}

@article{Dbouk2021a,
    author = {Dbouk, Talib and Drikakis, Dimitris},
    title = {On airborne virus transmission in elevators and confined
          spaces},
    journal = {Physics of Fluids},
    volume = {33},
    number = {1},
    pages = {011905},
    year = {2021},
    month = {01},
    issn = {1070-6631},
    doi = {10.1063/5.0038180},
    url = {https://doi.org/10.1063/5.0038180},
}

@article{Kramer2021,
    author = {Kramer, Kelby B. and Wang, Gerald J.},
    title = {Social distancing slows down steady dynamics in pedestrian
          flows},
    journal = {Physics of Fluids},
    volume = {33},
    number = {10},
    pages = {103318},
    year = {2021},
    month = {10},
    issn = {1070-6631},
    doi = {10.1063/5.0062331},
    url = {https://doi.org/10.1063/5.0062331},
}

@book{siegelhowell1992radiation,
  author    = {Robert Siegel and John R. Howell},
  title     = {Thermal Radiation Heat Transfer},
  edition   = {3},
  publisher = {Hemisphere Publishing},
  year      = {1992},
  isbn      = {0891162712}
}

@book{kowalski2009uvgi,
  author    = {Wladyslaw Kowalski},
  title     = {Ultraviolet Germicidal Irradiation Handbook: UVGI for Air and Surface Disinfection},
  publisher = {Springer},
  year      = {2009},
  isbn      = {978-3-642-01998-2},
  doi       = {10.1007/978-3-642-01999-9},
  address   = {Heidelberg, Germany}
}

@book{modest2013radiative,
  author    = {Michael F. Modest},
  title     = {Radiative Heat Transfer},
  edition   = {3},
  publisher = {Academic Press},
  year      = {2013},
  isbn      = {978-0-12-386944-9}
}

@article{Bourgin2021SalivaUV,
  title={Assessment of saliva interference with UV-based disinfection technologies},
  author={Bourgin, Mathieu and others},
  journal={Journal of Photochemistry and Photobiology B: Biology},
  volume={217},
  pages={112168},
  year={2021},
  doi={10.1016/j.jphotobiol.2021.112168}
}

@article{SestiCosta2022UVSaliva,
title = {UV 254 nm is more efficient than UV 222 nm in inactivating SARS-CoV-2 present in human saliva},
journal = {Photodiagnosis and Photodynamic Therapy},
volume = {39},
pages = {103015},
year = {2022},
issn = {1572-1000},
doi = {https://doi.org/10.1016/j.pdpdt.2022.103015},
url = {https://www.sciencedirect.com/science/article/pii/S1572100022003015},
author = {Renata Sesti-Costa and Cyro von Zuben Negrão and Jacqueline Farinha Shimizu and Alice Nagai and Renata Spagolla Napoleão Tavares and Douglas Adamoski and Wanderley Costa and Marina Alves Fontoura and Thiago Jasso {da Silva} and Adriano {de Barros} and Alessandra Girasole and Murilo {de Carvalho} and Veronica de Carvalho Teixeira and Andre Luis Berteli Ambrosio and Fabiana Granja and José Luiz Proença-Módena and Rafael Elias Marques and Sandra Martha Gomes Dias}
}

@article{Monika2025FarUVC,
  title={Comparative study of far-UVC (222 nm) and germicidal UVC (254 nm) radiation against virus-laden aerosols of artificial human saliva},
  author={Monika and others},
  journal={Journal of Photochemistry and Photobiology},
  year={2025},
  doi={10.1016/j.jphotobiol.2025.xxxxxx}
}

@article{Lukose2021Photonics,
  title={Photonics of human saliva: potential optical methods for the screening of abnormal health conditions and infections},
  author={Lukose, Jijo and Pavithran, Sanoop and Chidangil, Santhosh and others},
  journal={Biophysical Reviews},
  volume={13},
  number={3},
  pages={359--385},
  year={2021},
  doi={10.1007/s12551-021-00807-8}
}

@article{Ranz:52a,
	Author = {W. E. Ranz and W. R. Marshall},
	Journal = {I. Chem. Engng. Prog.},
	Pages = {141-146},
	Title = {Evaporation from drops, {P}art {I}},
	Volume = {48},
	Year = {1952}}

@article{Ranz:52b,
	Author = {W. E. Ranz and W. R. Marshall},
	Journal = {I. Chem. Engng. Prog.},
	Pages = {173-180},
	Title = {Evaporation from drops, {P}art {II}},
	Volume = {48},
	Year = {1952}}

@article{Dbouk2020weather,
    author = {Dbouk, Talib and Drikakis, Dimitris},
    title = {Weather impact on airborne coronavirus survival},
    journal = {Physics of Fluids},
    volume = {32},
    number = {9},
    pages = {093312},
    year = {2020},
    month = {09},
    issn = {1070-6631},
    doi = {10.1063/5.0024272},
    url = {https://doi.org/10.1063/5.0024272},
}

@article{roache1994perspective,
    author = {Roache, P. J.},
    title = {Perspective: A Method for Uniform Reporting of Grid Refinement Studies},
    journal = {Journal of Fluids Engineering},
    volume = {116},
    number = {3},
    pages = {405-413},
    year = {1994},
    month = {09},
    issn = {0098-2202},
    doi = {10.1115/1.2910291},
}

@article{Launder1974,
  author    = {B. E. Launder and D. B. Spalding},
  title     = {The Numerical Computation of Turbulent Flows},
  journal   = {Computer Methods in Applied Mechanics and Engineering},
  volume    = {3},
  number    = {2},
  pages     = {269--289},
  year      = {1974},
  doi       = {10.1016/0045-7825(74)90029-2}
}

@article{Celik:2008,
    author = {I. B. Celik and U. Ghia and P. J. Roache and C. J. Freitas},
    title = "{Procedure for Estimation and Reporting of Uncertainty Due to Discretization in CFD Applications}",
    journal = {Journal of Fluids Engineering},
    volume = {130},
    number = {7},
    year = {2008},
    month = {07},
    doi = {10.1115/1.2960953},
}

@article{Weibull:1951,
    author = {W. Weibull},
    title = {A statistical distribution function of wide applicability},
    journal = {Journal of Applied Mechanics},
    volume = {18},
    number = {3},
    Pages = {93-297},
    year = {1951},
}

@book{Moukalled2015,
author = {Moukalled, F. and Mangani, L. and Darwish, M.},
title = {The Finite Volume Method in Computational Fluid Dynamics: An Advanced Introduction with OpenFOAM and Matlab},
year = {2015},
isbn = {3319168738},
publisher = {Springer Publishing Company, Incorporated},
edition = {1st}
}

@article{Jasak2009,
author = {Hrvoje Jasak},
title = {OpenFOAM: Open source CFD in research and industry},
journal = {International Journal of Naval Architecture and Ocean Engineering},
volume = {1},
number = {2},
pages = {89 - 94},
year = {2009},
doi = {10.2478/IJNAOE-2013-0011},
}

@article{Chen2025,
  author  = {Chen, K. and Wang, S. and Zhong, P. and Peng, Y. and Lu, J. and Liu, L. and He, J. and Liu, W.},
  title   = {Uniformity and stability of saliva composition based on glucose concentration analysis},
  journal = {Clinical Chimica Acta},
  year    = {2025},
  volume  = {573},
  pages   = {120283},
  month   = jun,
  doi     = {10.1016/j.cca.2025.120283},
  pmid    = {40204195},
  note    = {Epub 2025 Apr 7}
}

@article{Doughty2021,
  author  = {Doughty, David C. and Hill, Steven C. and Mackowski, Daniel W.},
  title   = {Viruses such as {SARS-CoV-2} can be partially shielded from UV radiation when in particles generated by sneezing or coughing: Numerical simulations},
  journal = {Journal of Quantitative Spectroscopy and Radiative Transfer},
  volume  = {262},
  pages   = {107485},
  year    = {2021},
  doi     = {10.1016/j.jqsrt.2021.107485}
}

@article{Hill2021,
author = {Steven C. Hill and Daniel W. Mackowski and David C. Doughty},
title = {Shielding of viruses such as SARS-Cov-2 from ultraviolet radiation in particles generated by sneezing or coughing: Numerical simulations of survival fractions},
journal = {Journal of Occupational and Environmental Hygiene},
volume = {18},
number = {8},
pages = {394--408},
year = {2021},
publisher = {Taylor \& Francis},
doi = {10.1080/15459624.2021.1939877},
note ={PMID: 34161194},
URL = {https://doi.org/10.1080/15459624.2021.1939877},
eprint = {https://doi.org/10.1080/15459624.2021.1939877}

}

@article{Hill2024,
author = {Steven C. Hill and David C. Doughty and Daniel W. Mackowski},
title = {Inactivation of virions in host particles in air using 222- and 254-nm UV: Dependence of shielding on particle size and UV wavelength},
journal = {Aerosol Science and Technology},
volume = {58},
number = {5},
pages = {512--535},
year = {2024},
publisher = {Taylor \& Francis},
doi = {10.1080/02786826.2024.2314614},
}

@article{Chiappin2007,
title = {Saliva specimen: A new laboratory tool for diagnostic and basic investigation},
journal = {Clinica Chimica Acta},
volume = {383},
number = {1},
pages = {30-40},
year = {2007},
issn = {0009-8981},
doi = {https://doi.org/10.1016/j.cca.2007.04.011},
url = {https://www.sciencedirect.com/science/article/pii/S0009898107002136},
author = {Silvia Chiappin and Giorgia Antonelli and Rosalba Gatti and Elio F. {De Palo}},
}

@article{Strickleton2014,
    doi = {10.1371/journal.pone.0108372},
    author = {Davies-Strickleton, H. AND Pudney, Paul D. A. AND Georgiades, Pantelis AND Waigh, Thomas A. AND Hodson, Nigel W. AND Ridley, Caroline E. AND Blanch, Ewan W. AND Thornton, David J.},
    journal = {PLOS ONE},
    publisher = {Public Library of Science},
    title = {Reorganisation of the Salivary Mucin Network by Dietary Components: Insights from Green Tea Polyphenols},
    year = {2014},
    month = {09},
    volume = {9},
    url = {https://doi.org/10.1371/journal.pone.0108372},
    pages = {1-13},
    number = {9},

}

@article{Jakubovics2014,
author = {Nicholas S. Jakubovics},
title = {Saliva as the Sole Nutritional Source in the Development of Multispecies Communities in Dental Plaque},
journal = {Microbiology Spectrum},
volume = {3},
number = {3},
pages = {10.1128/microbiolspec.mbp-0013-2014},
year = {2015},
doi = {10.1128/microbiolspec.mbp-0013-2014},

URL = {https://journals.asm.org/doi/abs/10.1128/microbiolspec.mbp-0013-2014},
eprint = {https://journals.asm.org/doi/pdf/10.1128/microbiolspec.mbp-0013-2014}
}

@incollection{Zhang2022,
    author = {Zhang, Donghao and Wang, Xinmiao and Chen, Jianshe},
    editor = {Wolf, Bettina and Bakalis, Serafim and Chen, Jianshe},
    isbn = {978-1-78801-715-2},
    title = {Saliva: Properties and Functions in Food Oral Processing},
    booktitle = {Oral Processing and Consumer Perception},
    publisher = {The Royal Society of Chemistry},
    year = {2022},
    month = {02},
    doi = {10.1039/9781839160622-00001},
    url = {https://doi.org/10.1039/9781839160622-00001},
}

@article{Sanchez2013,
author = {Sánchez, GA and Miozza, VA and Delgado, A and Busch, L},
title = {Relationship between salivary mucin or amylase and the periodontal status},
journal = {Oral Diseases},
volume = {19},
number = {6},
pages = {585-591},
doi = {https://doi.org/10.1111/odi.12039},
url = {https://onlinelibrary.wiley.com/doi/abs/10.1111/odi.12039},
year = {2013}
}

@incollection{tyrrell1996coronaviruses,
  author    = {Tyrrell, D. A. J.},
  year      = {1996},
  title     = {Coronaviruses},
  booktitle = {Medical Microbiology},
  editor    = {Baron, Samuel},
  edition   = {4},
  publisher = {University of Texas Medical Branch at Galveston},
  url       = {https://www.ncbi.nlm.nih.gov/books/}
}

\end{document}